\documentclass[10pt]{article}
\pdfoutput=1
\usepackage{amstext,enumerate,amssymb,amsmath,bbm}
\usepackage{esint}
\usepackage{slashed}
\usepackage{framed}
\usepackage{float}
\usepackage{setspace}
\usepackage{todonotes}
\usepackage{cite}

\newcommand{\id}{{\mathbbm{1}}}

\newcommand{\n}{\nabla}

\newcommand{\z}{\zeta}
\newcommand{\zt}{\tilde{\z}}

\newtheorem{Pb}{Problem}

\newtheorem{Th}{Theorem}  
\newtheorem{Prop}{Proposition}  
  
\newtheorem{Cor}{Corollary}  
\newtheorem{Lem}{Lemma}  
\newtheorem{Def}{Definition} 
\newcommand{\bP}{\begin{Pb}\ \ } 
\newcommand{\eP}{\end{Pb}}  
\newcommand{\bt}{\begin{Th}\ \ }  
\newcommand{\et}{\end{Th}}  
\newcommand{\bp}{\begin{Prop}\ \ }  
\newcommand{\ep}{\end{Prop}}  
\newcommand{\bc}{\begin{Cor}\ \ }  
\newcommand{\ec}{\end{Cor}}  
\newcommand{\bl}{\begin{Lem}\ \ }  
\newcommand{\el}{\end{Lem}}  
\newcommand{\bd}{\begin{Def}\ \ }  
\newcommand{\ed}{\end{Def}} 

\numberwithin{equation}{section}

\begin{document}

 \rightline{LTH-1033} 
\vskip 1.5 true cm  
\begin{center}  
{\large Nernst branes from special geometry}\\[.5em]
\vskip 1.0 true cm   
{P.~Dempster$^{1,2}$, D.~Errington$^1$ and T.~Mohaupt$^{1}$} \\[3pt] 
$^1${Department of Mathematical Sciences\\ 
University of Liverpool\\
Peach Street \\
Liverpool L69 7ZL, UK\\  
P.Dempster@liv.ac.uk, D.Errington@liv.ac.uk, Thomas.Mohaupt@liv.ac.uk \\[1em]  
$^2${School of Physics \& Astronomy and Center for Theoretical Physics\\
Seoul National University\\
Seoul 151-747, KOREA \\
}
}
\vspace{5mm}
January 30, 2015. Revised: February 27, 2015.
\end{center}  
\vskip 1.0 true cm  
\baselineskip=18pt  
\begin{abstract}  
\noindent  
We construct new black brane solutions in $U(1)$ gauged ${\cal N}=2$
supergravity with a general cubic prepotential, which have 
entropy density $s\sim T^{1/3}$ as $T \rightarrow 0$ and thus satisfy the Nernst Law. 
By using the real formulation of special geometry, we are able
to obtain analytical solutions in closed form as functions
of two parameters, the temperature $T$ and the chemical potential
$\mu$. Our solutions
interpolate between hyperscaling violating Lifshitz
geometries with $(z,\theta)=(0,2)$ at the horizon and 
$(z,\theta)=(1,-1)$ at infinity. 
In the zero temperature limit, where the entropy density 
goes to zero, we recover the 
extremal Nernst branes of Barisch \textit{et al}, and the parameters
of the 
near horizon geometry change to $(z,\theta)=(3,1)$.

\end{abstract}


\newpage
 
\tableofcontents


\section{Introduction}

One of the most celebrated successes of string theory is the  AdS/CFT correspondence \cite{Maldacena}. 
This generates a powerful duality between asymptotically AdS gravitational theories and conformal field theories on the AdS boundary,
which is the simplest and best-studied example of the more general notion of a `gauge-gravity duality'. 
As a strong-weak coupling duality, the correspondence allows for the translation of non-perturbative field theory calculations into more tractable, perturbative calculations in gravity and vice-versa. 
This has enabled the exploration of previously inaccessible regimes of theoretical physics. Indeed, there are many examples of strongly coupled systems in condensed matter physics and it is hoped that  gauge-gravity duality may allow for a better understanding of these. Significant progress has already been made in this direction, leading to the development of the AdS/CMT correspondence (see \cite{Hartnoll:2009sz,Hartnoll:2011fn} and references therein).
Further recent progress has been to extend the correspondence to spacetimes which are not asymptotically-AdS but rather exhibit  hyperscaling violating and Lifshitz (hvLif) behaviour \cite{Dong,Sachdev:fermi},
thus extending the dictionary between gravity and 
condensed matter systems living on the boundary. 

The central idea in gauge/gravity duality is that each state in the bulk has a corresponding state in the dual field theory.
 In particular, black objects are dual to thermal ensembles in the field theory with the same thermodynamic properties (temperature, entropy, chemical potential, etc.) as the bulk spacetime~\cite{Witten:1998qj,Witten:1998zw}.

A natural starting point for the correspondence is to look at charged 
(Reissner-Nordstr\"om or `RN') extremal black holes and black branes
in AdS \cite{Chamblin:1999tk}.
However, like their asymptotically flat 
`cousins' they have a large non-zero entropy at zero temperature,
thus violating the 
Third Law of Thermodynamics, which states in its strictest version
that the entropy of a system should 
vanish in the zero temperature limit \cite{landau2013statistical}. 
While a non-vanishing entropy for certain classes of
extremal black holes is consistent with microstate counting 
for the corresponding
D-brane configuration in string theory 
\cite{Strominger:1996sh,Maldacena:1997de}, 
this still begs the question of whether 
one can find other gravitational systems which have
a zero entropy or entropy density at zero temperature. 
Apart from being an interesting
question about gravity, such systems are relevant for possible dualities
between gravity and condensed matter systems.

We remark that although `Nernst Law' is in the following used synonymously
with `Third Law of Thermodynamics', Nernst's original formulation
only requires that the difference in entropy between two
equilibrium states  related through a change in external
parameters goes to zero at zero temperature. This formulation 
is equivalent to the `process version' of the Third Law, which 
states that zero temperature cannot be reached by any physical
process in a finite number of steps. A process version of the
third law of black hole mechanics was already established
in \cite{Israel:1986}. However, the Nernst version or, equivalently,
the process version of the Third Law does not imply by itself the
slightly stronger version of the Third Law, due to Planck, which
states that the entropy itself goes to zero at zero temperature. 
This stricter version corresponds to systems with a unique
ground state, and thus is the generic situation in condensed
matter, although there is an extended debate about possible 
exceptions in specific systems, see for example 
\cite{Wald:1997qp,DHoker:2009bc,Hartnoll:2009sz,Hartnoll:2011fn}.

In the following we will be concerned with the explicit construction
of families of 
gravitational solutions which have zero entropy (or 
entropy density) in the extremal limit. Following conventions
in the literature, we will refer to the Third Law in its
stricter, Planckian, version as the Nernst Law.

Extremal brane solutions with 
vanishing entropy density at zero temperature have recently been
studied for a variety of bulk theories 
\cite{Goldstein:2009cv, Goldstein:2010aw, Gauntlett:2009dn,Horowitz:2009ij,DHoker:2009bc,Barisch:2011ui}  
and could have important applications in extending the dictionary
between condensed matter and gravity. 
They have been dubbed `Nernst branes' in \cite{Barisch:2011ui}, 
and it is believed that 
the corresponding non-extremal solutions exist and satisfy the Nernst 
Law, that is, these non-extremal solutions 
have a finite entropy which goes to zero when the temperature 
goes to zero while external parameters are kept fixed.
Finding such non-extremal solutions is important, since extremal
Nernst branes are not completely regular solutions. While all
curvature invariants remain finite at the horizon, tidal forces
become infinite and scalar fields take infinite values, which
suggests a breakdown of the underlying effective field theory
\cite{Hartnoll:2009sz,Barisch:2011ui}. A first step in addressing
this issue is to find non-extremal solutions, which 
can then be studied in the near extremal limit. In this context it
is clearly desirable to have completely explicit, analytical solutions.
However most results in the literature have to rely on a mixture
of analytical and numerical methods. Of course tidal forces may still
get very large at the horizon when one approaches the extremal limit
\cite{Horowitz:1997uc}, but analytical solutions will enable one to identify the
region in parameter space where the solution can be trusted and
possibly be mapped to condensed matter systems.

The second step in controlling the near horizon low temperature
behaviour is to embed 
the theory under consideration 
into a UV-complete theory, for which string theory
and its non-perturbative extension M-theory are arguably the best
candidates. In the low-energy limit the relevant stringy gravitational
backgrounds can be described in terms of supergravity. 
We will be
working in a set-up which can be described by $\mathcal{N}=2$ $U(1)$ gauged
supergravity with an arbitrary number of vector multiplets. 
Theories with $\mathcal{N}=2$ supersymmetry are natural
generalisations of the Einstein-Maxwell-Scalar theories underlying
dilatonic black hole and black brane solutions which have been 
studied extensively as potential duals of strongly coupled electron
systems \cite{Hartnoll:2009sz,Hartnoll:2011fn}. They
have the advantage that one can often
find exact, analytical answers, despite the fact that the couplings 
are not fixed by the matter content (as is the case for $\mathcal{N} \geq 4$
supersymmetry), but depend on arbitrary functions of the scalar fields,
which are subject to quantum and stringy corrections. 
While we do not discuss the string theory or M-theory
embedding explicitly, note that such theories
arise through heterotic flux compactifications on $K3\times T^2$ and
type-II flux compactifications on Calabi-Yau three-folds. We will 
not need to choose a specific model, and 
only assume that the vector multiplet couplings take 
the most general form that arises when working to leading order
in the Regge parameter $\alpha'$, and within the validity of string perturbation theory.
In other words, we only assume that the prepotential, which 
encodes the vector multiplet couplings, is of the so-called
very special type reviewed below. By working in a gauged supergravity
theory obtainable by flux compactification from string theory we will
have the option to further address the issues related to singularities
in the extremal limit at a later stage. For BPS black holes with 
vanishing entropy it is known that the inclusion of stringy 
higher curvature corrections in supergravity 
\cite{LopesCardoso:1998wt,LopesCardoso:2000qm}
leads to regular solutions with 
finite entropy \cite{Dabholkar:2004dq}, and the entropy function
formalism demonstrates that this mechanism is robust and does not
depend on supersymmetry and details of the higher curvature corrections
\cite{Sen:2005wa}. We refer to 
\cite{Hartnoll:2009sz,Hartnoll:2011fn,DHoker:2009bc} for a further
discussion of the possible implications of quantum and string corrections
to the zero temperature behaviour and the `fate' of the Nernst Law.

 Within the framework of four-dimensional $\mathcal{N}=2$ $U(1)$ gauged supergravity coupled to vector multiplets, 
 extremal Nernst branes have previously been constructed in \cite{Barisch:2011ui} using a first-order rewriting of the equations of motion, and by considering
a specific model: the so-called STU-model. 
 However a similar rewriting for their non-extremal counterparts has so far proven elusive, and the only known examples~\cite{heatup} have been constructed by deforming the metric of the corresponding five-dimensional extremal solution 
\cite{BarischDick:2012gj}
and imposing suitable consistency conditions.
 In this paper we are able to provide a systematic construction of non-extremal Nernst branes by directly solving the second-order equations of motion.
Moreover, our results will not only apply to a particular model, but to 
all models where the prepotential is of the very special type.
This gain in generality and systematics should help to expand 
the AdS/CMT dictionary considerably in the future.

We now present a brief overview of the results in this paper.
We start with a theory of $n$ $\mathcal{N}=2$ vector multiplets coupled to $U(1)$ gauged supergravity, with prepotential
\[
F(X)=\frac{f(X^1,\ldots ,X^n)}{X^0},
\]
where $f$ is homogeneous of degree three. If $f$ is a homogeneous 
\textit{polynomial} of degree three (which is not required for our methods
to apply), then the corresponding theory can be obtained
by dimensional reduction from five dimensions. Moreover, such prepotentials
capture perturbative string effects to leading order in $\alpha'$ if the 
model can be embedded into heterotic or type-II string theory. In this
case the supergravity lift to five dimensions becomes a lift from 
type-II string theory to M-theory.

Within these models we restrict ourselves to static black brane solutions.
Apart from this we will impose that the scalar fields take
purely imaginary values, as for such  `axion-free' field configurations 
there is a systematic simplification of the equations of motion.
Since we impose stationarity in four dimensions, 
we can perform a time-like dimensional reduction to obtain an effective 
three-dimensional Euclidean theory.
The degrees of freedom in three dimensions can then be repackaged using the real formulation of special geometry developed in \cite{Mohaupt:cmap},
 which has been used to construct solutions to both gauged \cite{Klemm:2012yg,Klemm:2012vm,Gnecchi:2013mja} and ungauged \cite{staticaxfree} theories of supergravity coupled to vector multiplets.

Since our ability to obtain explicit non-extremal solutions depends on
using a specific formalism, let us briefly summarize the underlying 
principles without going into technical details. 
\begin{itemize}
\item
Instead of using the physical four-dimensional scalar fields $z^A$, 
$A=1, \ldots, n$, we work on the `big moduli space' parametrized
by scalar fields $X^I$, $I=0,\ldots, n$. The additional (complex)
degree of freedom is compensated for by a local $\mathbbm{C}^*$ gauge
symmetry. Working on the big moduli space has the advantage that the
number of scalar fields and gauge fields matches.
\item
We use the real formulation of special K\"ahler geometry, which 
replaces the complex scalars $X^I$ by real scalars $q^a$, 
$a=0, \ldots, 2n+1$ and
which replaces the holomorphic prepotential $F(X^I)$ by a real
Hesse potential $H(q^a)$. This leads to a simpler, and fully
covariant, behaviour of all relevant quantities under 
electric-magnetic duality. 
\item
Upon dimensional reduction, the Kaluza-Klein scalar $\phi$ 
is absorbed into the real scalars $q^a$, which results in 
the `radial' direction of the big moduli space becoming 
a physical (rather than gauge) degree of freedom. \\
We postpone fixing the remaining $U(1) \subset \mathbbm{C}^*$ 
gauge symmetry to preserve electric-magnetic duality. The
resulting three-dimensional theory depends on $4n+5$
real scalars $q^a, \hat{q}^a, \tilde{\phi}$, subject to one
local gauge symmetry, where $\hat{q}^a$ are dual to the four-dimensional
gauge fields and $\tilde{\phi}$ is dual to the Kaluza-Klein vector.
While $q^a, \hat{q}^a$ are vectors under electric-magnetic duality,
$\tilde{\phi}$ is a scalar.
\item
We impose an ansatz which corresponds, from the four-dimensional
point of view, to a static solution with purely imaginary 
scalar fields. This determines $\tilde{\phi}$ and half of the
fields $q^a, \hat{q}^a$ in terms of the remaining fields, and
also fixes the residual $U(1)$ gauge symmetry. By abuse of 
notation, we denote the remaining independent fields 
by $q^a, \hat{q}^a$ (with a restricted range of $a$, depending
on the precise version of the ansatz). 
\item
When we now proceed to solve 
the time-reduced three-dimensional
equations of motion, their particular
structure allows us to 
obtain solutions in closed form. 
\end{itemize}

We remark that while some of the above ingredients are well known 
to people working on $\mathcal{N}=2$ supergravity, it is critical 
that these elements are put together into a systematic structure.
The key element that we use and preserve is electric-magnetic duality, which
acts on the fields by symplectic transformations.\footnote{We refer the
reader to \cite{deWit:2001pz,Cardoso:2012nh} for a comprehensive review of 
electric-magnetic duality in supergravity.} 
Our choice of
variables, which all transform as symplectic tensors, leads to the 
simplifications and systematics that we exploit. We observe that this
works despite the fact that the electric-magnetic duality group is broken to a 
discrete subgroup thereof by the presence of gauging (a scalar potential),
and despite the fact that our ansatz restricts us from the full symplectic
group to a subgroup.


Solving the three-dimensional equations of motion directly results in an instanton solution depending on a number of integration constants, which are \textit{a priori} undetermined.
However, in order that this solution lifts to a \textit{regular} black brane in four dimensions we have to impose suitable regularity conditions. In particular, we require that the four-dimensional solution has a finite entropy density, which happens to simultaneously ensure that the scalar fields take finite values on the horizon.
For a given set of charges and fluxes, we are then left with a two-parameter family of black brane solutions parametrised by a temperature $T$ and chemical potential $\mu$, which can both be freely varied.
In the limit of zero temperature, we recover the extremal Nernst branes of \cite{Barisch:2011ui}.
Therefore we interpret our solutions as non-extremal (or `hot') Nernst branes.
Indeed, it turns out that the entropy density goes to zero as $T\rightarrow 0$ for fixed charges/fluxes, in agreement with the Nernst Law.
Our solutions interpolate between hyperscaling violating Lifshitz
geometries with $(z,\theta)=(0,2)$ at the horizon and 
$(z,\theta)=(1,-1)$ at infinity, where $z$ is the dynamical critical 
exponent and where $\theta$ is the hyperscaling violating exponent.
In the zero temperature limit the
near horizon geometry changes to $(z,\theta)=(3,1)$.


This paper is organised as follows.
In Section~\ref{SecRealForm} we review the real formulation of special geometry as applied to $\mathcal{N}=2$ $U(1)$ gauged supergravity with both electric and magnetic fluxes, relegating the more technical details to the appendices.
We then reduce this theory over a time-like direction and determine the equations of motion of the three-dimensional theory for general static field configurations, before concentrating on the case of purely imaginary field configurations.
In Section~\ref{stusec} we solve the aforementioned equations of motion for the case where we have a single electric charge and some number of electric fluxes.
Having found a solution to the three-dimensional equations of motion we then lift it back to a four-dimensional solution and determine the conditions imposed on the various integration constants by regularity, before carrying out an analysis of the properties of the solution.
In Section~\ref{SecMagBrane} we apply our method to the case where we instead switch on a single magnetic charge and a single magnetic flux, whilst keeping $(n-1)$ of the electric fluxes.
Section~\ref{SecDiscussion} contains our conclusions. We also include
a brief initial discussion of our results 
in the context of holography.

\section{$\mathcal{N}=2$ gauged supergravity and the real formulation of special geometry}\label{SecRealForm}

\subsection{Lagrangian of $\mathcal{N}=2$ $U(1)$ gauged supergravity}

We begin with the well-known bosonic Lagrangian of $\mathcal{N}=2$ Fayet-Iliopoulos $U(1) \subset SU(2)_R$ gauged supergravity coupled to $n$ vector multiplets.
Our conventions follow those of~\cite{Klemm:2012vm, Mohaupt:cmap}\footnote{Note that in e.g.\ \cite{staticaxfree}, the opposite sign was used for the Einstein-Hilbert term of the corresponding ungauged theory, which leads to some sign-flips compared to the Einstein equations presented there.}
\begin{equation}\label{4dLag}
e_4^{-1} \mathcal{L}_4 =-\frac{1}{2} Y R_{(4)} - g_{IJ} \partial_{\hat{\mu}} X^I \partial^{\hat{\mu}} \bar{X}^J + \frac{1}{4} \mathcal{I}_{IJ} F^I_{\hat{\mu} \hat{\nu}} F^{J| \hat{\mu} \hat{\nu}} + \frac{1}{4} \mathcal{R}_{IJ} F^I_{\hat{\mu} \hat{\nu}} \tilde{F}^{J| \hat{\mu} \hat{\nu}} - V \left( X, \bar{X} \right),
\end{equation}
where $\hat{\mu},\hat{\nu}=0,\dots,3$ are four-dimensional spacetime indices, and $I,J=0, \dots, n$ label the four-dimensional gauge fields: $n$ from the vector multiplets and one, the graviphoton, from the gravity multiplet. 
For later convenience we use a formulation of the theory which
contains $n+1$ complex scalar fields $X^I$ which are subject to local
dilatations and $U(1)$ transformations. The $n$ physical scalars remaining
after gauge fixing can be parametrised as $z^A = X^A/X^0$, where
$A=1,\ldots, n$. While the physical scalars $z^A$ parametrise a
projective special K\"ahler (PSK) manifold, the $X^I$ parametrise
a conic affine special K\"ahler (CASK) manifold, which is a complex
cone over the PSK manifold. Conversely, the PSK manifold can be obtained
as the K\"ahler quotient of the CASK manifold with respect to the
$\mathbbm{C}^*$-action generated by dilatations and $U(1)$ transformations.
In physical terms this quotient amounts to gauge fixing the local
$\mathbbm{C}^*$ action, as discussed below. All terms in (\ref{4dLag})
except the scalar potential $V(X,\bar{X})$ are completely determined 
by the holomorphic prepotential $F(X^I)$, which is homogeneous of degree 2.
Prior to gauge fixing dilatations, the space-time Ricci scalar, $R_4$, is
multiplied by the conformal compensator
\[
Y = - i (X^I \bar{F}_I - F_I \bar{X}^I) ,
\]
where derivatives of the prepotential are denoted
$F_I = \frac{\partial F}{\partial X^I}$, etc. The tensor
\[
g_{IJ} = - \frac{\partial^2}{\partial X^I \partial \bar{X}^J} \log Y ,
\]
is the horizontal lift of the physical (PSK) scalar metric to the
CASK manifold. It has a two-dimensional kernel which reflects the fact
that the $X^I$ only represent $n$ complex physical degrees of freedom. 
The vector couplings are given by
\[
{\cal N}_{IJ} = {\cal R}_{IJ} + i {\cal I}_{IJ} =
\bar{F}_{IJ} + i \frac{N_{IK} X^K N_{JL} X^L}{ - X^M N_{MN} X^N}\;,
\]
where $N_{IJ} = 2 \mbox{Im} F_{IJ}$. 

We now turn to the $\mathbbm{C}^*$ gauge fixing. 
Dilatations are fixed by imposing the D-gauge
\begin{equation}\label{dgaugex}
-i\left(X^I\bar{F}_I -F_I\bar{X}^I\right)= \kappa^{-2} \;,
\end{equation}
which in particular brings the Einstein-Hilbert term in 
(\ref{4dLag}) to the standard form $-\frac{1}{2\kappa^2}R_4$.
Likewise $U(1)$ transformations can be fixed by imposing any condition transverse to the $U(1)$ action, such as $\text{Im} \left( X^0 \right)=0$.
However, as discussed in more detail in~\cite{Mohaupt:cmap,staticaxfree}, 
it is often advantageous to postpone $U(1)$ gauge fixing until reducing
the theory and starting to solve the resulting equations of motion. 
In particular,
upon imposing the D-gauge \eqref{dgaugex} one has 
\[
g_{IJ} \partial_{\hat{\mu}} X^I \partial^{\hat{\mu}} \bar{X}^J =
\bar{g}_{AB} \partial_{\hat{\mu}} z^A \partial^{\hat{\mu}} \bar{z}^B ,
\]
where $\bar{g}_{AB}$ is the positive definite (PSK) metric of the
physical scalars $z^A$. Working with the scalars $X^I$ has the advantage
that we retain covariance under symplectic transformations, and will
result in a more convenient form of the equations of motion after
reduction. Note that while the D-gauge removes one real degree of
freedom from the $X^I$, the second unphysical degree of freedom
is taken care of by the remaining local $U(1)$ symmetry, see \cite{Mohaupt:cmap}
for details. Geometrically, imposing the D-gauge while keeping the local
$U(1)$ symmetry corresponds to working on a $U(1)$ principal bundle over
the PSK manifold.

The four-dimensional Lagrangian \eqref{4dLag} also includes a 
scalar potential $V(X, \bar{X})$, which as in~\cite{Barisch:2011ui} 
is given as
\begin{equation}\label{scalpot}
V(X,\bar{X}) = N^{IJ} \partial_I W \partial_J \bar{W} - 2\kappa^2 | W |^2,
\end{equation}
with a superpotential of the form
\begin{equation}\label{superpot}
W=2\left(g^I F_I - g_I X^I\right),
\end{equation}
where $g^I,g_I$ parametrize the $U(1)$ gauging. Since superpotentials
of the form (\ref{superpot}) arise in flux compactifications, we refer
to them as magnetic and electric fluxes, respectively.
Note that we have included an explicit factor of $\kappa^2$ in 
\eqref{scalpot} using dimensional analysis. We will use this later
to rewrite the potential in terms of real variables.
For reference, 
we note that the $X^I$ have mass dimension $-1$ while the flux parameters
have dimension $-2$, so that $W$ has dimension $-3$. Since $N_{IJ}$ and,
hence, its inverse $N^{IJ}$ are homogeneous of degree 0, they have
dimension 0, and $V$ has dimension $-4$, as required. 
We also remark that for later 
convenience we have 
re-scaled the flux parameters by a factor of 2 relative to 
\cite{Barisch:2011ui}. Moreover, we have not factorized the flux
parameters into a dimensionful coupling and dimensionless parameters,
but kept them dimensionful.

\subsection{Reduction to three dimensions}

Imposing that the background is stationary, so that all of the fields are independent of time, we can reduce the four-dimensional action \eqref{4dLag} over a time-like direction in order to obtain an effective three-dimensional Euclidean action.
We decompose the four-dimensional metric as
\begin{equation}\label{4dmetric1}
ds^2_4 =-e^{\phi}\left(dt +V_\mu dx^\mu\right)^2 +e^{-\phi}ds^2_3,
\end{equation}
where $\phi$ and $V_\mu$ are the Kaluza-Klein scalar and vector respectively, and we have left the three-dimensional part of the metric undetermined for now.
Following the procedure for time-like dimensional reduction outlined in \cite{Mohaupt:cmap}, and noting that the scalar potential remains unchanged throughout the reduction process, one obtains the three-dimensional Lagrangian \cite{Klemm:2012vm}
\begin{eqnarray}\label{3dLagr1}
e_3^{-1}\mathcal{L}_3 &=& -\frac{1}{2}R_{(3)} 
-\tilde{H}_{ab}\left(\partial_\mu q^a\partial^\mu q^b -\partial_\mu \hat{q}^a \partial^\mu \hat{q}^b\right) +\frac{1}{2H}V \nonumber \\
& & -\frac{1}{H^2}(q^a\Omega_{ab}\partial_\mu q^b)^2
+\frac{2}{H^2}(q^a\Omega_{ab}\partial_\mu \hat{q}^b)^2 \nonumber \\
& & -\frac{1}{4H^2}(\partial_\mu\tilde{\phi}+2\hat{q}^a\Omega_{ab}\partial_\mu \hat{q}^b)^2 .
\end{eqnarray}
We have written all of the three-dimensional degrees of freedom using the conventions of the real formulation of special geometry developed in \cite{Mohaupt:cmap}, and afterwards set $\kappa=1$ for the remainder of the paper. While we give a brief summary here, more details can be found
 in Appendix~\ref{appendix:realspecial}. 
The three-dimensional action contains $4n+5$ scalar fields $(q^a, \hat{q}^a, 
\tilde{\phi})$ which are subject to one local $U(1)$ symmetry and
hence has $4n+4$ independent scalar degrees of freedom. 
While the $q^a$ combine the four-dimensional scalars $z^A$ with the
Kaluza-Klein scalar $\phi$, the $\hat{q}^a$ contain the degrees of
freedom of the four-dimensional gauge fields, and $\tilde{\phi}$
is dual to the Kaluza-Klein vector. 
The constant tensor
\[\Omega_{ab} = \left( \begin{array}{cc}
0 & \mathbbm{1} \\
- \mathbbm{1} & 0 \\
\end{array} \right) 
\] 
is the symplectic form of the CASK manifold expressed in real variables
$q^a$. 
The tensor $\tilde{H}_{ab}$ is given by
\[
\tilde{H}_{ab} = \frac{\partial^2 \tilde{H}}{\partial q^a \partial q^b}\;,\;\;\;
\tilde{H} = - \frac{1}{2} \log (-2H) \;,
\]
where the Hesse potential $H$ is related to the prepotential $F$ 
by the Legendre transformation given in~\eqref{legendre}.

As shown in the appendices, the scalar potential $V$ is given in terms of the real coordinates as 
\begin{equation}\label{scalarpotreal}
\frac{1}{H} V(q) = - 2 g^ag^b \left[ \tilde{H}_{ab} -4q_aq_b-2 \frac{\left(\Omega q \right)_a \left( \Omega q \right)_b}{H^2} \right] \;,
\end{equation}
where the dual scalars $q_a$ are defined by $q_a = -\tilde{H}_{ab} q^b$.

Substituting this expression into \eqref{3dLagr1} the three-dimensional Lagrangian  becomes
\begin{eqnarray}\label{3dLagr2}
e_3^{-1}\mathcal{L}_3 &=& -\frac{1}{2}R_{(3)} 
-\tilde{H}_{ab}\left(\partial_\mu q^a\partial^\mu q^b -\partial_\mu \hat{q}^a \partial^\mu \hat{q}^b +g^a g^b\right) \nonumber \\
& & -\frac{1}{H^2}(q^a\Omega_{ab}\partial_\mu q^b)^2
+\frac{2}{H^2}(q^a\Omega_{ab}\partial_\mu \hat{q}^b)^2 \nonumber \\
& & +4(g^a q_a)^2 +\frac{2}{H^2}(q^a\Omega_{ab}g^b)^2 
-\frac{1}{4H^2}(\partial_\mu\tilde{\phi}+2\hat{q}^a\Omega_{ab}\partial_\mu \hat{q}^b)^2 .
\end{eqnarray}

In the following we will restrict ourselves to static solutions, i.e.\ set $V_\mu=0$ in \eqref{4dmetric1},
 for which the final term in \eqref{3dLagr2} vanishes \cite{Mohaupt:cmap}.
The equations of motion for $\hat{q}^a$ are then given by
\begin{equation}\label{qhateom1}
\n_\mu \left(\tilde{H}_{ab}\partial^\mu \hat{q}^b\right)
+2\n_\mu\left(\frac{1}{H^2}q^b\Omega_{ba}(q^c\Omega_{cd}\partial^\mu \hat{q}^d)\right)=0,
\end{equation}
whilst those for $q^a$ read
\begin{align}\label{qeom1}
& \n_\mu \left(\tilde{H}_{ab}\partial^\mu q^b\right) -\frac{1}{2}\partial_a\tilde{H}_{bc}\left(\partial_\mu q^b\partial^\mu q^c -\partial_\mu \hat{q}^b \partial^\mu \hat{q}^c +g^b g^c\right) \nonumber \\
& -\frac{1}{2}\partial_a\left(\frac{1}{H^2}\right)(q^b\Omega_{bc}\partial_\mu q^c)^2
+\n_\mu \left(\frac{1}{H^2}q^b\Omega_{ba}(q^c\Omega_{cd}\partial^\mu q^d)\right)
-\frac{1}{H^2}\Omega_{ab}\partial_\mu q^b(q^c\Omega_{cd}\partial^\mu q^d) \nonumber \\
& +\partial_a\left(\frac{1}{H^2}\right)(q^c\Omega_{cd}\partial^\mu \hat{q}^d)^2
+\frac{2}{H^2}\Omega_{ab}\partial_\mu \hat{q}^b (q^c\Omega_{cd}\partial^\mu \hat{q}^d) \nonumber \\
& +4\tilde{H}_{ab}g^b (g^c q_c) 
+\partial_a\left(\frac{1}{H^2}\right)(q^b\Omega_{bc}g^c)^2
+\frac{2}{H^2}\Omega_{ab}g^b (q^c\Omega_{cd}g^d) =0.
\end{align}
Finally, the three-dimensional Einstein equations are
\begin{align}\label{3dEE1}
& -\frac{1}{2}R_{(3)\mu\nu}
-\tilde{H}_{ab}\left(\partial_\mu q^a\partial_\nu q^b -\partial_\mu \hat{q}^a \partial_\nu \hat{q}^b\right)
-\frac{1}{H^2}(q^a\Omega_{ab}\partial_\mu q^b)(q^c\Omega_{cd}\partial_\nu q^d) \nonumber \\
& +\frac{2}{H^2}(q^a\Omega_{ab}\partial_\mu \hat{q}^b)(q^c\Omega_{cd}\partial_\nu \hat{q}^d)
+g_{\mu\nu}\left(-\tilde{H}_{ab}g^a g^b +4(g^a q_a)^2 +\frac{2}{H^2}(g^a\Omega_{ab}q^b)^2\right) =0.
\end{align}

\subsection{Purely imaginary field configurations}\label{SSecPI}

We concentrate in this paper on purely imaginary (PI) field configurations,
which we define to be those for which the complex scalars\footnote{The scalars $Y^I$ are rescaled versions of the scalars $X^I$. See \eqref{rescale} for the definition.} $z^A=Y^A/Y^0$ are purely imaginary \cite{staticaxfree}.
Moreover, we restrict ourselves to a class of prepotentials of the form
\begin{equation}\label{genericprepotential}
F(Y)= \frac{f(Y^1,\ldots,Y^n)}{Y^0},
\end{equation}
where the function $f$ is homogeneous of degree three and real-valued when evaluated on real fields.
For the case of ungauged $\mathcal{N}=2$  supergravity, such models were considered in \cite{staticaxfree}.
Note that those models with  $f$ a cubic polynomial are precisely the `very special' prepotentials for  which the solutions can be uplifted to five dimensions.
Since we choose to fix the $U(1)$ gauge by taking $\mathrm{Im} Y^0=0$, this is equivalent to setting $x^A=\mathrm{Re}Y^A$ to zero. Models obtainable 
from five dimensions are invariant under constant shifts $x^A \rightarrow
x^A + C^A$, and, hence, PI configurations will be referred to as `axion-free'.

For the class of models \eqref{genericprepotential} the scalar fields $q^a$ take the form \cite{staticaxfree}
\[
\left.(q^a)\right|_{\mathrm{PI}} =(x^0,0,\ldots,0; 0, y_1,\ldots,y_n),
\]
and hence we see that $q^a\Omega_{ab}\partial_\mu q^b=0$. 
Following \cite{staticaxfree} we extend the PI condition to the scalars $\hat{q}^a$ by imposing
\[
\left.(\partial_\mu \hat{q}^a)\right|_{\mathrm{PI}} =\frac{1}{2}(\partial_\mu \z^0,0,\ldots, 0; 0,\partial_\mu \zt_1,\ldots, \partial_\mu \zt_n ),
\]
which sets also $q^a\Omega_{ab}\partial_\mu \hat{q}^b=0$. 
The quantities $\partial_\mu \zeta^I$ and $\partial_\mu \tilde{\zeta}_I$
encode the four-dimensional field strengths, see (\ref{fieldstrengths}).
 
In the same way, we extend the PI condition to the fluxes $g^a$ by imposing
\[
\left.(g^a)\right|_{\mathrm{PI}} =(g^0,0,\ldots,0; 0,g_1,\ldots,g_n),
\] 
which sets $q^a\Omega_{ab}g^b=0$.

We then find that the equations of motion \eqref{qhateom1}--\eqref{qeom1} and the three-dimensional Einstein equations \eqref{3dEE1} simplify to
\begin{equation}\label{qhateom2}
\n_\mu \left(\tilde{H}_{ab}\partial^\mu \hat{q}^b\right) =0, 
\end{equation}
\begin{equation}\label{qeom2}
 \n_\mu \left(\tilde{H}_{ab}\partial^\mu q^b\right) -\frac{1}{2}\partial_a\tilde{H}_{bc}\left(\partial_\mu q^b\partial^\mu q^c -\partial_\mu \hat{q}^b \partial^\mu \hat{q}^c +g^b g^c\right)+4\tilde{H}_{ab}g^b (g^c q_c) 
=0 ,
\end{equation}
and
\begin{equation}\label{3dEE2}
-\frac{1}{2}R_{(3)\mu\nu}
-\tilde{H}_{ab}\left(\partial_\mu q^a\partial_\nu q^b -\partial_\mu \hat{q}^a \partial_\nu \hat{q}^b\right)
+g_{\mu\nu}\left(-\tilde{H}_{ab}g^a g^b +4(g^a q_a)^2 \right) =0.
\end{equation}

It turns out to be useful to write the equations of motion in terms of the dual variables $q_a$ and $\hat{q}_a$ defined in Appendix \ref{appendix:realspecial}.
In terms of these, the equations \eqref{qhateom2}--\eqref{3dEE2} become
\begin{equation}\label{qhateom3}
\n^2 \hat{q}_a =0,
\end{equation}
\begin{equation}\label{qeom3}
\n^2 q_a +\frac{1}{2}\partial_a\tilde{H}^{bc}\left(\partial_\mu q_b\partial^\mu q_c -\partial_\mu \hat{q}_b\partial^\mu \hat{q}_c\right) -\frac{1}{2}\partial_a\tilde{H}_{bc}g^b g^c +4\tilde{H}_{ab}g^b (g^c q_c)=0,
\end{equation}
and
\begin{equation}\label{3dEE3}
-\frac{1}{2}R_{(3)\mu\nu}
-\tilde{H}^{ab}\left(\partial_\mu q_a\partial_\nu q_b -\partial_\mu \hat{q}_a \partial_\nu \hat{q}_b\right)
+g_{\mu\nu}\left(-\tilde{H}_{ab}g^a g^b +4(g^a q_a)^2 \right) =0.
\end{equation}

In the next section we will look for solutions of \eqref{qhateom3}--\eqref{3dEE3} which can be lifted to regular non-extremal black branes in four dimensions.


\section{Non-extremal black branes}\label{stusec}

Our aim in this section is to construct a family of non-extremal black branes in the $\mathcal{N}=2$ gauged supergravity theory \eqref{4dLag} with prepotential \eqref{genericprepotential}.
Restricting our attention to the PI configurations described in Section \ref{SSecPI}, it can be shown that the Hesse potential takes the 
form\cite{staticaxfree}
\begin{equation}\label{Hessepot}
H=-\frac{1}{4}\left(-q_0f(q_1,\ldots,q_n)\right)^{-\frac{1}{2}} .
\end{equation}
For general functions $f$, the form of the metric $\tilde{H}^{ab}$ is fairly complicated \cite{staticaxfree}.
However, since the field $q_0$ decouples from the rest, we can compute
\[
\tilde{H}^{00}=\frac{1}{4q_0^2}, \quad q^0=-\frac{1}{4q_0}, 
\]
and this will be sufficient to find solutions valid for any 
choice of $f$.
We remark here upon a slight abuse of notation which we will make throughout the remainder of this paper.
Specifically, we denote by $q_A$ with $A=1,\ldots,n$ those scalar fields which are actually the $(A+n+1)$'th components of the vector $(q_a)$. 
The same is true of the components $\tilde{H}_{AB}$ of the metric, which should properly be the $(A+n+1,B+n+1)$ components of $\tilde{H}_{ab}$. This notation
is convenient since $(q^0,q_A)$ are the remaining non-trivial $q^a$-fields 
within our ansatz.

For simplicity 
we will concentrate on solutions which are supported by a single electric charge $Q_0$ and electric fluxes $g_1,\ldots, g_n$ in this section.
However, as we will see in Section \ref{SecMagBrane}, the methods introduced in 
the following
can be easily extended to deal also with solutions with a single magnetic charge switched on and sourced by both electric and magnetic fluxes.
The systematic investigation of dyonic black branes will be carried out in a future publication \cite{futurepaper}.

\subsection{Einstein equations}

We make a brane-like ansatz for  the three-dimensional metric:
\begin{equation}\label{3dmetric}
ds^2_3 =e^{4\psi}d\tau^2 +e^{2\psi}(dx^2+dy^2),
\end{equation}
where $\psi=\psi(\tau)$ is some function to be determined.
This form of the metric can always be obtained from the more commonly used
form $ds^2_{3}=dr^2 +e^{2\psi}(dx^2 +dy^2)$ by a suitable redefinition
$r\rightarrow \tau$.
We also impose that all fields $q_a$ and $\hat{q}_a$ depend only on $\tau$.
The coordinate $\tau$ has been chosen such that it is an affine
parameter for geodesic curves on the scalar target space parametrized
by $q^a$ and $\hat{q}^a$. Equivalently, the
$\tau$-dependent part of the three-dimensional Laplace operator is given by 
$\frac{\partial^2}{\partial \tau^2}$.

The non-zero components of the Ricci tensor are given by
\[
R_{\tau\tau}=2\ddot{\psi}-2\dot{\psi}^2, \quad
R_{xx}=R_{yy}=e^{-2\psi}\ddot{\psi},
\]
where the dot denotes differentiation with respect to $\tau$.
With this choice the three-dimensional Einstein equations \eqref{3dEE3} become
\begin{equation}\label{xxEE}
-\tilde{H}_{ab}g^ag^b +4(q_a g^a)^2 -\frac{1}{2}e^{-4\psi}\ddot{\psi}=0,
\end{equation}
for $\mu=\nu\neq\tau$ and
\begin{equation}\label{HamConst}
\tilde{H}^{ab}\left(\dot{q}_a\dot{q}_b -\dot{\hat{q}}_a\dot{\hat{q}}_b\right) =\dot{\psi}^2 -\frac{1}{2}\ddot{\psi},
\end{equation}
for $\mu=\nu=\tau$, where we have used \eqref{xxEE}. 
Equation \eqref{HamConst} is the Hamiltonian constraint which needs to
be imposed on solutions $(q_a(\tau), \hat{q}_a(\tau))$
of the second order scalar field equations. We remark that since
we have consistently reduced the full field equations, we do not
need to impose this constraint by hand, but have retained it as
a field equation following from an action principle.

\subsection{Scalar equations of motion}

We now turn to the equations of motion for the fields $q_a$ and $\hat{q}_a$. 
We start with the $\hat{q}_a$ equations of motion, which read simply
\[
\ddot{\hat{q}}_a =0, 
\]
and can be integrated once to find
\begin{equation}\label{qhatsoln}
\dot{\hat{q}}_a =K_a,
\end{equation}
for some constants $K_a$,
which are proportional to the electric and magnetic charges of the solution, $K_a=(-Q_I,P^I)$ \cite{staticaxfree}. 
The explicit relations between the $\hat{q}_a$ and the field strengths can be found in Appendix~\ref{appendix:realspecial}.
For the case at hand we only have a single electric charge $Q_0$,
and so the only non-zero component of $\dot{\hat{q}}_a$ is $\dot{\hat{q}}_0=-Q_0$.


We turn now to the $q_a$ equations of motion \eqref{qeom3}, which become
\begin{equation}\label{qeom4}
e^{-4\psi}\ddot{q}_a +\frac{1}{2}\partial_a\tilde{H}^{bc} e^{-4\psi}\left(\dot{q}_b\dot{q}_c -\dot{\hat{q}}_b\dot{\hat{q}}_c\right)
-\frac{1}{2}\partial_a\tilde{H}_{bc}g^bg^c +4\tilde{H}_{ab}g^b (q_c g^c)=0.
\end{equation}

For the models \eqref{genericprepotential} without magnetic flux, $g^0 =0$, 
on which we concentrate in this section, 
the $q_0$ equation of motion decouples from the others.
Indeed, using \eqref{qhatsoln} with $K_0=-Q_0$ the $q_0$ equation of motion becomes
\begin{equation}
\ddot{q}_0 -\frac{\dot{q}_0^2-Q_0^2}{q_0}=0 .
\end{equation}
This takes precisely the same form as in the ungauged case \cite{staticaxfree} and can be solved with
\begin{equation}\label{q0soln}
q_0(\tau) =\pm -\frac{Q_0}{B_0}\sinh \left(B_0\tau +B_0 \frac{h_0}{Q_0}\right), 
\end{equation}
for some constants $B_0$ and $h_0$.
Since the solution \eqref{q0soln} is invariant under $B_0\rightarrow -B_0$, we can take $B_0\geq 0$ without loss of generality.
It will turn out that $B_0$ acts as a non-extremality parameter for the full solution.
Furthermore, as we will see later explicitly, $\tau$ naturally takes
values $0 \leq \tau < \infty$. Thus 
in order that $q_0\neq 0$ for $\tau\geq 0$ we will have to require $\mathrm{sign}(h_0)=\mathrm{sign}(Q_0)$.


The $q_A$ equations of motion, for $A=1,\ldots,n$, become\footnote{We choose to leave the sum explicit here for convenience.}
\begin{equation}\label{qAeom1}
e^{-4\psi}\ddot{q}_A +\frac{1}{2}e^{-4\psi}\sum_{B,C=1}^n \partial_A\tilde{H}^{BC} \dot{q}_B \dot{q}_C
-\frac{1}{2}\sum_{B,C=1}^n (\partial_A\tilde{H}_{BC})	g_B g_C 
 +4\sum_{B=1}^n \tilde{H}_{AB} g_B\left(\sum_{C=1}^n q_Cg_C\right) =0. 
\end{equation}
Multiplying by $q^A$ and summing over $A$ gives
\begin{equation}\label{qsumeom1}
e^{-4\psi}\sum_{A=1}^n q^A \ddot{q}_A +e^{-4\psi}\sum_{A,B=1}^n \tilde{H}^{AB}\dot{q}_A \dot{q}_B
+\sum_{A,B=1}^n \tilde{H}_{AB}\,g_A g_B -4\left(\sum_{A=1}^n g_Aq_A\right)^2 =0,
\end{equation}
where we have made use of the homogeneity properties of the metric $\tilde{H}_{ab}$, viz.\ $q^a\partial_a\tilde{H}^{bc}=2\tilde{H}^{bc}$ and $q^a\partial_a\tilde{H}_{bc}= -2\tilde{H}_{bc}$.
One can now compare this equation to \eqref{xxEE}, which for the model at hand becomes
\[
-\sum_{A,B=1}^n \tilde{H}_{AB}\,g_A g_B +4\left(\sum_{A=1}^n g_Aq_A\right)^2 -\frac{1}{2}e^{-4\psi}\ddot{\psi}=0 \;.
\]
Substituting from this into the last two terms of \eqref{qsumeom1} we obtain
\begin{equation}
\sum_{A=1}^n  q^A \ddot{q}_A +\sum_{A,B=1}^n \tilde{H}^{AB}\dot{q}_A \dot{q}_B =\frac{1}{2}\ddot{\psi}.
\end{equation}
The left-hand side of this equation can be rewritten as a total derivative
\[
\sum_{A=1}^n  q^A \ddot{q}_A +\sum_{A,B=1}^n \tilde{H}^{AB}\dot{q}_A \dot{q}_B =\frac{d}{d\tau} \left(\sum_{A=1}^n q^A \dot{q}_A\right),
\]
and so we can integrate to find
\begin{equation}\label{qsumeom2}
\sum_{A=1}^n q^A \dot{q}_A =\frac{1}{2}\dot{\psi} -\frac{1}{4}a_0,
\end{equation}
for some integration constant $a_0$, where we have chosen the factor for later convenience.
Now, using the identity $\partial^a\tilde{H}=\tilde{H}^{ab}q_b$ \cite{staticaxfree} one can show furthermore that
\[
\frac{d\tilde{H}}{d\tau} =-q^0\dot{q}_0 -\sum_{A=1}^n q^A \dot{q}_A =\frac{\dot{q}_0}{4q_0}- \sum_{A=1}^n q^A \dot{q}_A .
\]
Substituting this expression into \eqref{qsumeom2} and further integrating gives
\[
-2\psi +a_0\tau +b_0 =4\tilde{H} -\log(-q_0) =-2\log\left(-4H\cdot (-q_0)^{1/2}\right),
\]
where we have used the definition of $\tilde{H}$ given in \eqref{Htildedef}, and have chosen the definition of the integration constant $b_0$ for later convenience.
Substituting the explicit expression for the Hesse potential \eqref{Hessepot}  we therefore find
\begin{equation}\label{qpsisoln}
\log\left(f(q_1,\ldots,q_n)\right)=-2\psi +a_0\tau +b_0.
\end{equation}

Let us now return to the Hamiltonian constraint \eqref{HamConst} which, upon substituting the expression \eqref{q0soln}, becomes
\begin{equation}\label{HamConst2}
\sum_{A,B=1}^n \tilde{H}^{AB}\dot{q}_A\dot{q}_B=\dot{\psi}^2 -\frac{1}{2}\ddot{\psi}-\frac{1}{4}	B_0^2 .
\end{equation}


So far we have the following picture: 
the equations of motion for the $q_A$ are given by the set of coupled equations  \eqref{qAeom1}.
The solutions $q_A(\tau)$ of \eqref{qAeom1} should then satisfy the two constraints \eqref{qpsisoln} and \eqref{HamConst2}.

We proceed by imposing that the $q_A$ are all proportional,
which will in turn mean that all of the physical scalar fields $z^A$ are proportional to one another\footnote{
Of course, it would be interesting for future work to investigate whether solutions can be found, for generic choices of the flux parameters, where this assumption is relaxed.}.
 Specifically, we set $q_A(\tau)=\xi_A q(\tau)$ for some constants $\xi_A$.
In terms of this ansatz, the constraints \eqref{HamConst2} and (the derivative of) \eqref{qpsisoln} become
\begin{equation}\label{qconstraints}
3\left(\frac{\dot{q}}{q}\right)^2 =4\dot{\psi}^2 -2\ddot{\psi} -B_0^2, \quad 3\left(\frac{\dot{q}}{q}\right)=-2\dot{\psi}+a_0 .
\end{equation}
We have made use here of the homogeneity properties of $f$ and the metric $\tilde{H}^{ab}$, as well as the identity $\tilde{H}_{ab}(q)q^a q^b=1$ \cite{Mohaupt:cmap} which implies, for the models at hand, that
\[
\sum_{A,B=1}^n \tilde{H}^{AB}(\xi)\xi_A\xi_B =\frac{3}{4}.
\]

The two equations \eqref{qconstraints} can be combined into a second-order non-linear differential equation for $\psi(\tau)$:
\begin{equation}\label{psiDE}
\ddot{\psi} -\frac{4}{3}\dot{\psi}^2 -\frac{2}{3}a_0\dot{\psi} +\frac{1}{2}B_0^2 +\frac{1}{6}a_0^2 =0.
\end{equation}
Introducing the variable
\[
y\equiv \exp\left(-\frac{4}{3}\psi-\frac{1}{3}a_0\tau\right),
\]
this becomes
\[
\ddot{y}-\omega^2 y=0, 
\]
for
\[
\omega^2 =\frac{2}{3}B_0^2 +\frac{1}{3}a_0^2,
\]
and hence can be solved by
\begin{equation}\label{psisoln}
\exp\left(-\frac{4}{3}\psi-\frac{1}{3}a_0\tau\right)
=\frac{\alpha}{\omega}\sinh\left(\omega\tau +\omega\beta\right),
\end{equation}
where $\alpha$ and $\beta$ are integration constants, and we have taken  $\omega$ to be the positive root without loss of generality.
 Note that the right hand side should be non-negative for all $\tau>0$,
and hence we should pick $\alpha>0$ and $\beta\geq 0$.
The solution \eqref{psisoln} now determines the function $\psi(\tau)$ appearing in the metric ansatz in terms of some integration constants, which we will fix in Section~\ref{SSecphysicality}.


We can now use \eqref{psisoln} to find an expression for $q(\tau)$.
Indeed, differentiating \eqref{psisoln} with respect to $\tau$ and substituting into the second equation in \eqref{qconstraints} we obtain
\[
\frac{\dot{q}}{q} =\frac{1}{2}\omega\coth(\omega\tau+\omega\beta) +\frac{1}{2}a_0.
\]
This can be integrated up to find
\begin{equation}
q(\tau) =\Lambda e^{\frac{1}{2}a_0\tau}\left(\sinh(\omega\tau+\omega\beta)\right)^{\frac{1}{2}},
\end{equation}
where $\Lambda$ is an integration constant.
Since we have set all of the $q_A$ proportional to each other, we can therefore write
\[
q_A=\lambda_A e^{\frac{1}{2}a_0\tau}\left(\sinh(\omega\tau+\omega\beta)\right)^{\frac{1}{2}},
\]
for some constants $\xi_A\equiv\lambda_A/\Lambda$.
Substituting this into \eqref{qAeom1} we find that $q_1g_1=\ldots=q_n g_n$, and
that the $q_A$ equation of motion is satisfied provided the integration constants $\lambda_A$ are related to the electric fluxes $g_A$ via
\[
\lambda_A=\pm \frac{3}{8n g_A}\left(\frac{\alpha^3}{\omega}\right)^{\frac{1}{2}}.
\]

Returning to \eqref{qpsisoln} then determines the constant $b_0$ in terms of $\alpha$ and the fluxes $g_A$ as
\[
e^{b_0} =\pm \left(\frac{3\alpha}{8n}\right)^3 f\left(\frac{1}{g_1},\ldots,\frac{1}{g_n}\right).
\]
Finally, the Kaluza-Klein scalar $\phi$ appearing in the metric ansatz \eqref{4dmetric1} is determined in terms of the $q_a$ via the $D$-gauge condition \eqref{Dgaugespecre} and the explicit form of the Hesse potential \eqref{Hessepot}.


To summarise, we find that the scalars $q_a$ are given by
\begin{eqnarray}
q_0 &=& \pm -\frac{Q_0}{B_0}\sinh\left(B_0\tau +B_0 \frac{h_0}{Q_0}\right), \label{q01} \\
q_A &=& \pm \frac{3}{8n g_A}\left(\frac{\alpha^3}{\omega}\right)^{\frac{1}{2}}e^{\frac{1}{2}a_0\tau} \left(\sinh(\omega\tau+\omega\beta)\right)^{\frac{1}{2}} \quad\mbox{for}\quad A=1,\ldots , n, \label{qA1}
\end{eqnarray}
whilst the metric degrees of freedom are given by
\begin{eqnarray}
e^{-4\psi} &=& \left(\frac{\alpha}{\omega}\right)^3 \sinh^3(\omega\tau +\omega\beta) e^{a_0\tau}, \label{psi1} \\
e^{\phi} &=& \frac{1}{2}(-q_0)^{-\frac{1}{2}}(f(q_1,\ldots, q_n))^{-\frac{1}{2}}. \label{phi1}
\end{eqnarray}
The $\pm$ signs in \eqref{q01}--\eqref{qA1} should be chosen such that the function $e^\phi$ is well-defined.

\subsection{The Nernst brane solution}\label{SSecphysicality}

In this section we want to look at the conditions on the various integration constants which give rise to regular black brane solutions in four dimensions.
In particular, we impose that
our solution has finite entropy density, which is the relevant regularity condition for solutions with non-compact horizon.

Let us recall the form of the four-dimensional metric in the $\tau$ coordinates:
\begin{equation}\label{4dmetric2}
ds^2_4 =-e^{\phi}dt^2 +e^{-\phi+4\psi}d\tau^2 +e^{-\phi+2\psi}(dx^2 +dy^2).
\end{equation}

We will see below that for a suitable choice of integration constants
$\tau = \infty$ is an event horizon, while
$\tau \rightarrow 0$ is the asymptotic regime at infinite distance.
The regularity of the solution within the bulk 
between horizon and infinity depends on the detailed properties of the 
function $f$. In particular, when evaluating $f$ on the solution,
we require that it has neither zeroes (so that there are 
in particular no changes of sign of $e^\phi$) nor poles. 
Given the experience with
similar issues for black hole solutions and domain walls, 
one expects that such solutions
exist for any prepotential arising in string theory upon suitable
restriction of the integration constants \cite{Kallosh:2000rn,Mayer:2003zk}. 
In any case, such questions
can only be investigated explicitly on a case-by-case basis, while
we restrict ourselves to questions that can be answered irrespective
of the choice of $f$.

The position of the event horizon can be found by looking at the value of $\tau$ for which the norm of the Killing vector field $k=\partial_t$ vanishes.
Since $k^2=g_{tt}=-e^{\phi}\sim \exp(-\frac{1}{2}B_0\tau -\frac{3}{4}a_0\tau -\frac{3}{4}\omega\tau)$ as $\tau\rightarrow\infty$, we can identify the horizon with the limiting value $\tau\rightarrow\infty$ provided $a_0\geq 0$.
If $a_0 <0$ then the position of the horizon will change depending on the relative magnitudes of $|a_0|$ and $B_0$, and so we will take $a_0\geq 0$ in what follows.

The area of the horizon is given by
\[
\int dx dy \left. e^{-\phi+2\psi}\right|_{\tau\rightarrow\infty} ,
\]
which is divergent since the $x$ and $y$ coordinates are non-compact.
However, we can still define a finite entropy density
$s$ provided the factor
$\left.e^{-\phi+2\psi}\right|_{\tau\rightarrow\infty}$ remains finite.
From the expressions \eqref{psi1}--\eqref{phi1} one can show that in this limit we have
\[
\left. e^{-\phi+2\psi}\right|_{\tau\rightarrow\infty} \sim \exp\left(\frac{1}{2}B_0\tau +\frac{1}{4}a_0\tau -\frac{3}{4}\omega\tau\right).
\]
In order that this be finite and non-zero at the horizon we therefore require
\[
\frac{1}{2}B_0 +\frac{1}{4}a_0=\frac{3}{4}\omega,
\]
which turns out to be equivalent to fixing
$a_0=B_0$.
Note that in this case we likewise have $\omega=B_0$.

We still at this stage have four integration constants $h_0, B_0, \alpha, \beta$ which are \textit{a priori}  yet to be determined.
However, note that we can always absorb $\beta$ into a shift of $\tau$ and a redefinition of the constants $\alpha$ and $h_0$.
Indeed, it will be useful to set $\beta=0$ at this stage so that the asymptotic  region of the solution is  at $\tau =0$. 
Moreover, we see that in the extremal $B_0\rightarrow 0$ limit, the expression \eqref{psisoln} becomes $e^{-4/3\psi}=\alpha\tau$. Hence, we can scale $\tau$ to set $\alpha=1$,  matching the conventions of the extremal Nernst brane of \cite{Barisch:2011ui}.
We are therefore left with a two-parameter family of solutions to the three-dimensional equations of motion, parametrised by $B_0$ and $h_0$, which we will interpret in terms of thermodynamic quantities in Section \ref{SSecProperties}.

Before moving on to study properties of the solution, we summarise the results so far:  the scalars $q_a$ and $\hat{q}_a$ are given by
\begin{eqnarray}
q_0 &=& \pm -\frac{Q_0}{B_0}\sinh\left(B_0\tau +B_0 \frac{h_0}{Q_0}\right), \label{q02} \\
q_A &=& \pm \frac{3}{8n g_A} B_0^{-\frac{1}{2}}e^{\frac{1}{2}B_0\tau} \left(\sinh(B_0\tau)\right)^{\frac{1}{2}} \quad\mbox{for}\quad A=1,\ldots ,n, \label{qA2} \\
\dot{\hat{q}}_0 &=& -Q_0,
\end{eqnarray}
whilst the metric degrees of freedom are given by
\begin{eqnarray}
e^{-4\psi} &=& \frac{1}{B_0^3} \sinh^3(B_0\tau) e^{B_0\tau}, \label{psi2} \\
e^{\phi} &=& \frac{1}{2}(-q_0)^{-\frac{1}{2}}(f(q_1,\ldots, q_n))^{-\frac{1}{2}}. \label{phi2}
\end{eqnarray}
The physical scalar fields $z^A=Y^A/Y^0$ can be determined from the 
expressions
\[
Y^A=-\frac{i}{2} e^{\phi}q_A, \quad Y^0=-\frac{1}{4q_0} ,
\]
which were obtained in~\cite{staticaxfree}, see 
Appendix \ref{appendix:realspecial}. 
We find
\begin{equation}
z^A =-i\left(\frac{-q_0 q_A^2}{f(q_1,\ldots,q_n)}\right)^{\frac{1}{2}} .
\end{equation}
Note that for $B_0\neq 0$, $q_0$ and $q_A$ all behave as $\exp(B_0\tau)$ when $\tau\rightarrow \infty$. We will show in the following section that this implies that  
the physical scalar fields take finite values on the horizon for $B_0\neq 0$. 

\subsection{Properties of the Nernst brane solution}\label{SSecProperties}

We now turn to an analysis of various properties of the solution obtained in Section~\ref{SSecphysicality}, postponing a fuller discussion to Section~\ref{SecDiscussion}.

\subsubsection*{A coordinate change}

It is convenient to  introduce the radial coordinate $\rho$ via
\[
e^{-2B_0\tau} =1-\frac{2B_0}{\rho}\equiv W(\rho).
\]
With this definition, the asymptotic region is situated at $\rho\rightarrow\infty$, while the horizon is at $\rho =2B_0$.
In terms of $\rho$, we find the expressions
\[
q_0 =\pm \frac{\mathcal{H}_0}{W^{1/2}} ,\quad\mbox{and}\quad
q_A =\pm \frac{3}{8n g_A}(\rho W)^{-1/2} \quad\mbox{for}\quad A=1,\ldots,n,
\]
where we have introduced the function\footnote{We follow the
sign conventions of \cite{staticaxfree}. See in particular Section 5.3.1
for a comparison of conventions for the $STU$-model.}
\[
\mathcal{H}_0 (\rho) =-\left[\frac{Q_0}{B_0}\sinh \left(\frac{B_0 h_0}{Q_0}\right) +\frac{Q_0 e^{-\frac{B_0 h_0}{Q_0}}}{\rho}\right] .
\]
The physical scalar fields $z^A(\rho)$ then take the form
\[
z^A =-i\left(\pm \frac{8n}{3 g_A^2}f\left(\frac{1}{g_1},\ldots,\frac{1}{g_n}\right)^{-1}\rho^{1/2}\mathcal{H}_0\right)^{\frac{1}{2}}.
\]
Hence, for $h_0\neq 0$ we find the asymptotic behaviour $z^A\sim \rho^{1/4}$, whilst for $h_0=0$ we find $z^A\sim \rho^{-1/4}$.

The four-dimensional line element \eqref{4dmetric2} becomes
\begin{equation}\label{4dmetricrho}
ds^2_4 = -\mathcal{H}^{-\frac{1}{2}}W\rho^{\frac{3}{4}} dt^2 
+\mathcal{H}^{\frac{1}{2}} \rho^{-\frac{7}{4}} \frac{d\rho^2}{W} +\mathcal{H}^{\frac{1}{2}} \rho^{\frac{3}{4}}(dx^2+dy^2) ,
\end{equation}
where we have found it convenient to define
\[
\mathcal{H}(\rho)\equiv \pm 4\left(\frac{3}{8n}\right)^3 f\left(\frac{1}{g_1},\ldots,\frac{1}{g_n}\right)\mathcal{H}_0(\rho).
\]
From this form of the metric, it is clear that the limit $B_0\rightarrow 0$ can be achieved simply by setting $W=1$ and
\[
\mathcal{H}_{0|\mathrm{ext}}=-\left(h_0 +\frac{Q_0}{\rho}\right).
\]
In this case we reproduce the extremal Nernst brane solutions of \cite{Barisch:2011ui}, albeit in different coordinates.
This identifies $B_0$ as a parameter encoding the non-extremality of the solution.

For $h_0=0$, the harmonic function for both the extremal and non-extremal solutions becomes $\mathcal{H}_0(\rho)=-Q_0/\rho$.
The line element \eqref{4dmetricrho} then becomes
\begin{equation}\label{4dmetrich0=0}
ds^2_{4|h_0=0} =-Z^{-\frac{1}{2}}W\rho^{\frac{5}{4}}dt^2 +Z^{\frac{1}{2}}\rho^{-\frac{9}{4}}\frac{d\rho^2}{W} +Z^{\frac{1}{2}}\rho^{\frac{1}{4}}(dx^2+dy^2) ,
\end{equation}
where we have defined
\[
Z\equiv \pm 4\left(\frac{3}{8n}\right)^3 Q_0 f\left(\frac{1}{g_1},\ldots,\frac{1}{g_n}\right) \;,
\]
with the sign chosen such that $Z$ is positive. 
The corresponding extremal solution can be obtained by setting the `blackening
factor' $W=1$ in \eqref{4dmetrich0=0}.

\subsubsection*{Near-horizon behaviour}

To investigate the near-horizon behaviour of the line element \eqref{4dmetricrho}, we  define $r^2\equiv \rho-2B_0$ and zoom in on the region $r\approx 0$.
We then find that for $B_0\neq 0$ the near-horizon metric looks like
\begin{eqnarray}\label{4dNHmetric}
ds^2_4 &=& -\left(Z e^{\frac{B_0 h_0}{Q_0}}\right)^{-1/2} (2B_0)^{1/4} r^2 dt^2 
+4 \left(Z e^{\frac{B_0 h_0}{Q_0}}\right)^{1/2} (2B_0)^{-5/4} dr^2 \nonumber \\
&& +\left(Z e^{\frac{B_0 h_0}{Q_0}}\right)^{1/2}(2B_0)^{1/4} (dx^2+dy^2),
\end{eqnarray}
which is the product of a two-dimensional Rindler spacetime with two-dimensional flat space. 
We also include, for comparison, the near-horizon behaviour of the extremal solution  which, after putting $\rho =R^{-4}$, becomes
\begin{equation}\label{NHExtmetric}
ds^2_{4|\textrm{Ext}} = \frac{1}{R}\left[ -\frac{1}{R^4} Z^{-\frac{1}{2}}dt^2 +16 Z^{\frac{1}{2}} dR^2 +Z^{\frac{1}{2}}(dx^2 +dy^2)\right] .
\end{equation}

By Wick rotating to Euclidean time $t\rightarrow t_E=it$ in \eqref{4dNHmetric} and enforcing regularity of the $t_E$ circle we can read off the temperature
\begin{equation}\label{temp}
4\pi T_H = Z^{-1/2} (2B_0)^{3/4}e^{-\frac{B_0 h_0}{2Q_0}} .
\end{equation}
We can also read off from \eqref{4dNHmetric} the entropy density of the solution, which is given by
\begin{equation}\label{entropy}
s=Z^{1/2}(2B_0)^{1/4}e^{\frac{B_0 h_0}{2Q_0}}.
\end{equation}
Note that from \eqref{temp} and \eqref{entropy} we can eliminate the integration constant $B_0$ in terms of the thermodynamic quantities $s$ and $T_H$ via.
\begin{equation}\label{B0eqn}
B_0 =2\pi s T_H.
\end{equation}

\subsubsection*{Asymptotic behaviour}

We now turn to a consideration of the asymptotic $\rho\rightarrow\infty$ properties of the line element \eqref{4dmetricrho}, which for $h_0\neq 0$ becomes
\[
ds^2_{4|\mathrm{asymp}} = \mathcal{H}(\infty)^{\frac{1}{2}}\rho^{\frac{1}{4}}
\left[ -\frac{1}{\mathcal{H}(\infty)}\rho^{\frac{1}{2}}dt^2 +\frac{d\rho^2}{\rho^2} +\rho^{\frac{1}{2}}(dx^2 +dy^2)\right].
\]
Note that this is the same for both the extremal and non-extremal solutions.
Making the coordinate change $\rho=R^{-4}$ then brings this to the form
\begin{equation}\label{4dasympmetric}
ds^2_{4|\mathrm{asymp}} = \frac{1}{R^3}\left[-\mathcal{H}(\infty)^{-\frac{1}{2}} dt^2 + 16\mathcal{H}(\infty)^{\frac{1}{2}} dR^2 +\mathcal{H}(\infty)^{\frac{1}{2}} (dx^2 +dy^2)\right] ,
\end{equation}
which is conformally $\mathrm{AdS}_4$ with boundary at $R=0$.

For the case $h_0=0$, the asymptotic limit corresponds simply to $W\rightarrow 1$ in \eqref{4dmetrich0=0}, from which we find the asymptotic line element  \eqref{NHExtmetric}, after a suitable coordinate redefinition.

\subsubsection*{Chemical potential}

The gauge field strength $F_{\tau t}^0$ is determined from the scalar field $\hat{q}^0$ via \eqref{fieldstrengths}:
\[
\dot{A}^0_t =2\dot{\hat{q}}^0 =2\tilde{H}^{00} \dot{\hat{q}}_0 = -\frac{Q_0}{2q_0^2}.
\]
Substituting in the expression \eqref{q02} and integrating with respect to $\tau$ gives
\begin{equation}\label{gaugefield}
A_t(\tau) =\frac{1}{2}\left(\frac{B_0}{Q_0}\right)\left[ \coth\left(B_0\tau +\frac{B_0 h_0}{Q_0}\right) -1\right],
\end{equation}
where we have chosen the integration constant such that $A_t(\infty)=0$, i.e.\ that the gauge fields vanish on the horizon\footnote{See e.g.\ \cite{Hartnoll:2009sz} for motivation for this condition.}.
The chemical potential $\mu$ is then given by the asymptotic value of $A_t$,
\begin{equation}\label{chempot}
\mu \equiv A_t(0) =\frac{1}{2}\left(\frac{B_0}{Q_0}\right)\left[ \coth\left(\frac{B_0 h_0}{Q_0}\right) -1\right] ,
\end{equation}
which diverges as $h_0\rightarrow 0$.
Note that in the extremal limit $B_0\rightarrow 0$ with $h_0\neq 0$ we get $\mu_{\mathrm{ext}}=1/(2h_0)$.

\subsubsection*{Thermodynamics and the Nernst Law}
We are now in a position to relate the integration constants $B_0$ and $h_0$ appearing in our solution to the thermodynamic quantities $s$, $T_H$ and $\mu$. 
In particular, we can rearrange \eqref{chempot} to find
\[
e^{\frac{2B_0 h_0}{Q_0}} =1+\frac{B_0}{Q_0\mu} =1+\frac{2\pi sT_H}{Q_0\mu} ,
\]
where we have used \eqref{B0eqn}.
Returning to \eqref{entropy} we then find an equation determining the entropy density as a function of the electric charge $Q_0$, fluxes $g_1,\ldots,g_n$, temperature $T_H$ and chemical potential $\mu$ of the black brane:
\begin{equation}\label{eqnofstate}
s^3=4\pi Z^2 T_H\left(1+\frac{2\pi s T_H}{Q_0\mu}\right).
\end{equation}
One consequence of \eqref{eqnofstate} is that, if we keep $Z$, $Q_0$ and $\mu$ fixed and send $T_H\rightarrow 0$, we see that $s\rightarrow 0$, which is precisely the strict (Planckian) formulation of the third law of thermodynamics \cite{landau2013statistical}.
This identifies the solution constructed in Section \ref{SSecphysicality} as a non-extremal (`hot') Nernst brane.

We can further analyse \eqref{eqnofstate} by looking at the dimensionless ratio $T_H/\mu$.
When $T_H/\mu$ is small, the second term in \eqref{eqnofstate} becomes negligible, and we find that the entropy density behaves as $s\sim T_H^{1/3}$.
On the other hand, when $T_H/\mu$ becomes large, the second term in \eqref{eqnofstate} dominates, and we find the behaviour $s\sim T_H$.

\begin{figure}[t]
\centering
\includegraphics[scale=0.5]{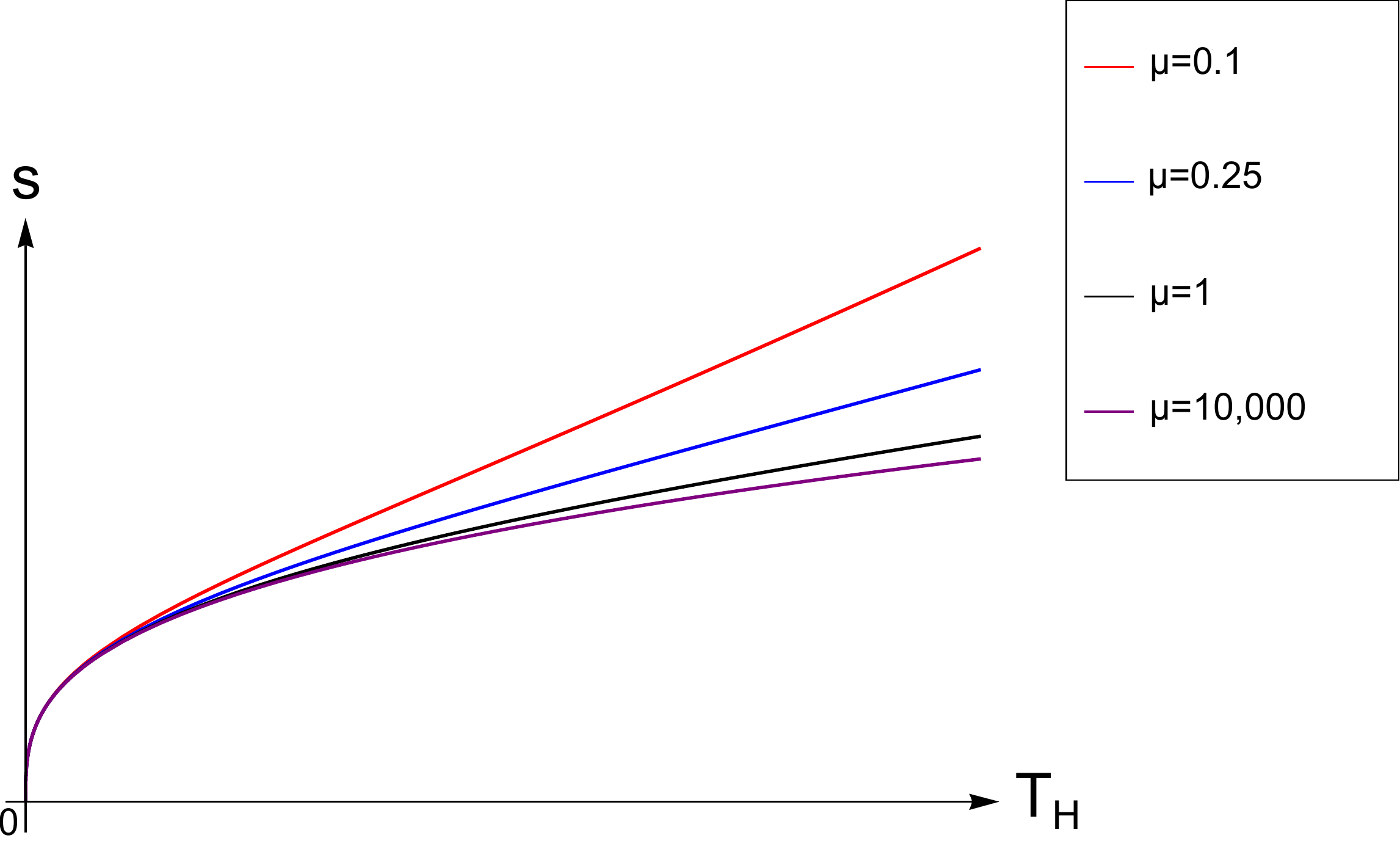}
\caption{Mathematica plot of \eqref{eqnofstate}, showing how entropy density $s$ varies with temperature $T_H$ for various values of the chemical potential $\mu$, and with $Q_0$ and $Z$ fixed.}
\label{stdiagram}
\end{figure}

In Figure \ref{stdiagram} we plot equation \eqref{eqnofstate} for various values of $\mu$, keeping $Q_0$ and $Z$ fixed.
This shows a) the Nernst Law behaviour $s\rightarrow 0$ as $T_H\rightarrow 0$, and b) the crossover from the behaviour  $s\sim T_H^{1/3}$ to $s\sim T_H$. 

\section{A magnetic black brane}\label{SecMagBrane}

We now turn our attention to a simple reformulation of the procedure in Section~\ref{stusec} which for a certain class of prepotentials allows us to construct non-extremal black branes carrying magnetic charge.
We will here simply present the supergravity solution, and leave a fuller discussion of the thermodynamics of magnetically-charged black branes for future work.

In particular, we are interested in prepotentials for which one of the fields $Y^1,\ldots,Y^n$ decouples from the others.
Without loss of generality, we can assume that $Y^1$ decouples, and consider prepotentials of the form
\[
F(Y)=\left(\frac{Y^1}{Y^0}\right)\tilde{f}(Y^2,\ldots,Y^n) ,
\]
where the function $\tilde{f}$ is homogeneous of degree 2.
This class is particularly interesting from the perspective of embedding the model into string theory as it contains  the tree-level heterotic prepotentials, which are linear in the heterotic dilaton $Y^1/Y^0$.
We consider black brane solutions  which are supported by a single magnetic charge $P^1$, a  magnetic flux $g^0$, and  electric fluxes $g_2,\ldots, g_n$.

In this case we see that the equations of motion can be solved in precisely the same way as in Section \ref{stusec},
with the field $q_1$ and magnetic charge $P^1$ playing the role of $q_0$ and $Q_0$ in the preceding section.
In particular, we have
\[
q_1 (\tau) = \pm \frac{P^1}{B_0}\sinh\left(B_0\tau +B_0\frac{h^1}{P^1}\right),
\]
whilst $q_0$ and $q_2,\ldots, q_n$ take the same form as \eqref{qA1} after replacing $g_1$ with $g^0$ in the obvious place. 
Moreover, the function $\psi$ remains unchanged and, since
\[
e^\phi =\frac{1}{2}(-q_0q_1 \tilde{f}(q_2,\ldots, q_n))^{-\frac{1}{2}},
\]
is symmetric in $q_0$ and $q_1$, we find that the line element takes the same form as in Section \ref{stusec}.
Looking at the near-horizon behaviour we again find that regularity of the solution imposes the same relation between the integration constants, $a_0=B_0$, as before.
The entropy density is therefore
\[
s=Z^{1/2}(2B_0)^{1/4}e^{\frac{B_0 h^1}{2 P^1}},
\]
whilst the temperature of the solution is given by
\[
4\pi T_H = Z^{-1/2}(2B_0)^{3/4}e^{-\frac{B_0 h^1}{2P^1}}.
\]

\section{Discussion and conclusions}\label{SecDiscussion}

In this paper we have provided a new technique for the construction of  non-extremal black brane solutions to large classes of  $\mathcal{N}=2$ $U(1)$ gauged supergravity models, utilising the techniques of time-like dimensional reduction followed by a rewriting of the effective three-dimensional degrees of freedom  through the real formulation of special geometry. 
In Section~\ref{stusec} we explicitly constructed a family of non-extremal black branes supported by a single electric charge and an arbitrary number of electric fluxes.
 This family of branes has an entropy density
behaving as $s\sim T^{1/3}$ for $T\rightarrow 0$, which therefore 
vanishes at $T=0$, where we recover the extremal
Nernst brane solutions of \cite{Barisch:2011ui}.
We anticipate that such non-extremal Nernst branes will have interesting applications in the context of holography, where they could prove useful in describing  dual field theory configurations at finite temperature and chemical potential which satisfy the Nernst Law. 

One issue with regards to a holographic interpretation is that our solutions do not fit naturally into the framework of AdS/CMT, since they do not asymptote to $\mathrm{AdS}_4$, but rather conformal $\mathrm{AdS}_4$, as seen in \eqref{4dasympmetric}.
Hence, the stress tensor of the dual field theory in the UV would not be scale invariant.
However, in recent years much progress has been made in understanding the holographic description of such `hyperscaling violating' theories, as well as the more general class of hyperscaling violating Lifshitz (hvLif) theories~\cite{Dong,Sachdev:fermi,perlmutter}, which we now review.

Consider spacetime geometries of the form (we use the conventions of~\cite{Dong})
\begin{equation}\label{hvlif}
ds_{d+2}^2 = r^{- \frac{2( d - \theta)}{d}} \left( -r^{-2 \left( z-1 \right)} dt^2 +  dr^2 + dx_i^2 \right),
\end{equation}
where $i=1,\ldots,d$ label the spatial directions on the boundary,
$z$ is the `dynamical critical' (Lifshitz) exponent, and $\theta$ is the `hyperscaling violating' exponent\footnote{We refer the reader to e.g.~\cite{Dong,Sachdev:fermi,perlmutter} for further details. For recent results on 
hvLif-like solutions in supergravity, see \cite{Bueno:2012sd,Bueno:2012vx}.}. 
Note that for $z=1$, $\theta=0$ one recovers the metric on $\mathrm{AdS}_{d+2}$. 

By looking at the near-horizon and boundary  behaviour of our solutions, we see that the Nernst brane interpolates between two hvLif geometries \eqref{hvlif} with $d=2$.
There are four cases of interest, corresponding to whether $h_0$ and $B_0$ are zero or non-zero:
\begin{itemize}
\item $h_0=0$, $B_0=0$: The solution becomes globally hvLif \eqref{NHExtmetric} with $(z,\theta)=(3,1)$. It has zero temperature and infinite chemical potential.
\item $h_0=0$, $B_0\neq 0$: The solution \eqref{4dmetrich0=0} has finite temperature and infinite chemical potential, and interpolates between a near-horizon Rindler geometry \eqref{4dNHmetric}, with $(z,\theta)=(0,2)$, and an asymptotic hvLif geometry with $(z,\theta)=(3,1)$.
\item $h_0\neq 0$, $B_0=0$: The solution has zero temperature and a finite chemical potential. It interpolates between a hvLif geometry with $(z,\theta)=(3,1)$ at the horizon, and the conformal $\mathrm{AdS}_4$ geometry~\eqref{4dasympmetric} with $(z,\theta)=(1,-1)$ at infinity.
This is the Nernst brane solution of \cite{Barisch:2011ui}.
\item $h_0\neq 0$, $B_0\neq 0$: The solution \eqref{4dmetricrho} has finite temperature and chemical potential, and interpolates between a near-horizon Rindler geometry with $(z,\theta)=(0,2)$ and the conformal $\mathrm{AdS}_4$ geometry with 
$(z,\theta)=(1,-1)$ at infinity.
\end{itemize}
Note that all of these values are consistent with the constraints imposed by the Null Energy Condition~\cite{Dong}.
We have therefore found, \textit{analytically}, a family of solutions which interpolate between two hvLif geometries.
This family is parametrised by the two integration constants $B_0$ and $h_0$, or equivalently by the temperature $T$ and chemical potential $\mu$ of the solution, both of which can be freely varied. Both parameters have a distinct effect
on the near horizon and asymptotic forms of the solution: while the extremal or 
zero temperature limit $B_0 \rightarrow 0$ changes the near horizon
solution from $(z,\theta)=(0,2)$ to $(z,\theta)=(3,1)$, the 
infinite chemical potential limit $h_0\rightarrow 0$ changes the
geometry at infinity from $(z,\theta)=(1,-1)$ to $(z,\theta)=(3,1)$.
If both limits are performed we obtain a global hvLif solution 
with $(z,\theta)=(3,1)$ which we interpret as the ground state of the given charge sector. Note that 
like any Lifshitz solution different from AdS it is not geodesically 
complete, and that the scalars are non-constant and run off to 
zero or infinity in the asymptotic regions. However, a similar 
behaviour can occur for domain wall solutions in gauged supergravity
which, for lack of more symmetric solutions, are interpreted as 
ground states. Sometimes this interpretation can be further justified
by an embedding into string theory or M-theory, see for example 
\cite{Mayer:2004sd}. While we leave studying the string theory 
embedding of our solutions for future work, we remark that the
interpretation is consistent with a limit where the temperature is zero 
and the chemical potential infinite.

Since so far solutions
interpolating between hvLif geometries have only been found by relying on
a mixture of analytical and numerical methods, we have made a significant
step forward, and expect that the techniques used and described in this
paper will be useful in making further progress.
While we leave searching for a concrete holographic dual of  the bulk geometries presented in this paper to future work, we can already make some interesting observations which shed some light on the properties which such a putative dual theory might possess.

Let us first consider the extremal ($B_0=0$) solution with $h_0=0$.
Since this is the gravitational ground state solution 
with $(z,\theta)=(3,1)$, zero temperature and infinite chemical 
potential, we expect it to be dual 
to the ground state of a $(2+1)$-dimensional QFT with hyperscaling exponent $\theta=1$ and Lifshitz exponent $z=3$. We remark that the specific value
$\theta=1$ for a QFT in $d=2$ space dimensions seems to be required for 
the description of states with hidden Fermi surfaces,
although a three-loop calculation gives $z=\frac{3}{2}$ rather than 
$z=3$ \cite{Sachdev:fermi}.

Now consider turning on some finite temperature $T>0$ on the field theory side.
By a simple scaling argument, one can argue \cite{Sachdev:fermi}
that the entropy density of the thermal state is related to the temperature as $s\sim T^{\frac{d-\theta}{z}}=T^{1/3}$. 
We therefore expect that the non-extremal Nernst brane with $h_0=0$ in \eqref{4dmetrich0=0} provides us with the relevant gravity dual to the $(2+1)$-dimensional QFT with $\theta=1$ and $z=3$ at finite temperature.
Indeed, taking $\mu\rightarrow\infty$
in the relation \eqref{eqnofstate}
we see that the entropy density of the brane solution is related to the temperature as $s\sim T^{1/3}$ which is the expected behaviour from the field theory arguments, and therefore consistent with our 
tentative interpretation.

\begin{figure}[t]
\centering
\includegraphics[scale=1.3,trim=5.5cm 9cm 6.5cm 10.5cm,clip=true]{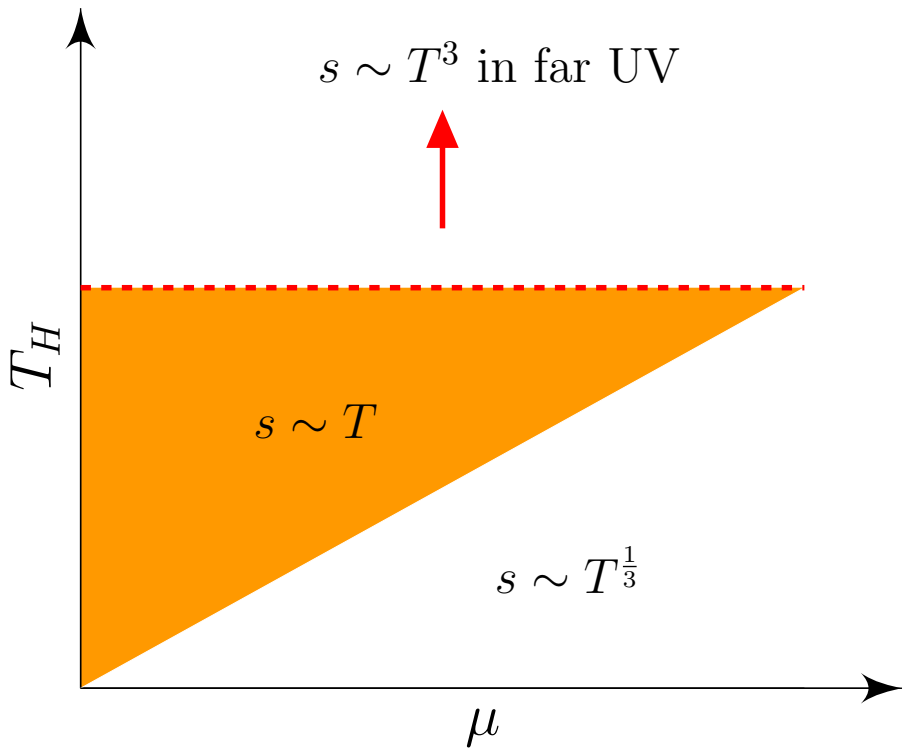}
\caption{The holographic phase diagram for our family of Nernst brane solutions in terms of horizon temperature, $T_H$, and chemical potential, $\mu$, which shows a smooth crossover between the two scaling regimes. We have also indicated that we anticipate a different scaling behaviour in the far UV where we don't expect that our supergravity solution accurately describes the tentative dual theory.}
\label{phasediagram}
\end{figure}

We now move on to consider what happens at finite chemical potential $\mu<\infty$, which corresponds to $h_0\neq 0$.
In this case, the extremal Nernst brane interpolates between a hvLif geometry with $(z,\theta)=(3,1)$ at the horizon, and a hvLif with $(z,\theta)=(1,-1)$ at infinity, which is conformal to $\mathrm{AdS}_4$.
One possible interpretation is as an RG flow between two QFTs:
one with hyperscaling exponent $\theta=-1$ in the UV; and one with hyperscaling exponent $\theta=1$ and Lifshitz exponent $z=3$ in the IR. 
As the gravity solution is smooth, and we do not seem to have a natural 
candidate for an order parameter identifying a phase transition, we think that
the more likely interpretation is that the UV `phase' and the IR `phase'
are related by smooth crossover. For the IR theory we expect that 
the entropy scales like $s\sim T^{\frac{d-\theta}{z}} = T^{\frac{1}{3}}$, 
which agrees with the behaviour of the Nernst brane solution for low
temperature $\frac{T}{\mu}\ll 1$. Adding temperature changes the
near horizon geometry, but leaves the asymptotic geometry at infinity
unchanged, which is consistent with interpreting these configurations
as thermal states. We therefore expect that the IR behaviour is 
correctly described by the Nernst brane solution, which in turn predicts
a scaling $s \sim T$ of the entropy for high temperatures, $\frac{T}{\mu}
\gg 1$. This however does not agree with the expected scaling of our 
tentative UV theory with $(z,\theta)=(1,-1)$, which predicts $s\sim T^3$.
We also note that the asymptotic UV geometry, while conformal to 
$\mathrm{AdS}_4$, cannot be interpreted as an alternative ground state of our
supergravity theory, because it is not, when taken as a global geometry,
part of our family of solutions. Moreover, the physical scalar
fields $z^A \sim \rho^{1/4}$ run off to infinity in the UV region,
which indicates strong coupling or decompactification. Taken together
this suggests that the description in terms of our
four-dimensional gauged supergravity theory is incomplete in the UV,
and that further degrees of freedom become relevant. If we accept
that the UV geometry correctly captures the thermodynamic behaviour 
then the corresponding UV theory should have a scaling behaviour
$s \sim T^3$ ($z=1, \theta=-1, d=2$). The resulting tentative
phase diagram is shown in Figure~\ref{phasediagram}.

The above mentioned analogy with domain walls together with the runaway
behaviour of the scalars suggests to interpret the UV behaviour as a
decompactification limit and to embed the four-dimensional 
supergravity theory into a higher dimensional theory. 
Since the class of prepotentials that we have considered in this paper 
includes those `very special' prepotentials for which the theory can be 
uplifted to five dimensions, the most obvious embedding is into five-dimensional
supergravity.
There are grounds to  believe~\cite{perlmutter} that the dimensional reduction 
of theories admitting $\mathrm{AdS}_D$ vacua would admit vacua with some 
nontrivial hvLif behaviour. Therefore we expect 
that  by lifting our solutions to five dimensions we will obtain
new asymptotically $\mathrm{AdS}_5$ finite temperature solutions in 
$\mathcal{N}=2$ gauged supergravity which still satisfy the Nernst 
Law\footnote{Although examples of such asymptotically $\mathrm{AdS}_5$ hot 
Nernst solutions were constructed in~\cite{heatup}, their solutions do not 
reduce to the finite temperature solutions presented here.}.
We will expand on this point in \cite{futurepaper5d}, and 
remark that an asymptotic $\mathrm{AdS}_5$ leads to a scaling of the 
entropy $s\sim T^{\frac{d-\theta}{z}} = T^3$, ($z=1, \theta=0, d=3$),
which is consistent with our proposed UV theory. 

We should also point out that there are issues
with the interpretation of our solutions if the temperature is strictly zero, 
since the Nernst brane solution has infinite tidal forces and run-away behaviour
of the scalars at the horizon in the extremal limit. This again 
indicates a breakdown of the effective description, and strictly speaking
the supergravity solution should only be trusted 
at low but finite temperature. 
Thus, as in the similar case of the holographic interpretation of 
hyperscaling violating solutions of Einstein-Maxwell-Dilaton theories
\cite{Dong}, the Nernst brane solution is not a valid description 
of its (tentative)
dual over the full range of the energy (radial coordinate)
from the UV (infinity) to the IR (horizon), but only over a finite
interval outside the horizon.
We leave it to future work to characterize the range of validity more
quantitatively, and to identify the necessary completions in the UV 
and IR using a string theory embedding. 
One possible strategy to further investigate the zero temperature limit
is to adapt formalisms that allow to 
include higher derivative terms. In ${\cal N}=2$ supergravity
a certain class of higher derivative terms (those encoded in the
so-called Weyl multiplet), which are related to the topological string, 
lead to a generalization of the framework of 
special geometry, on which we relied in the article 
\cite{LopesCardoso:1998wt,LopesCardoso:2000qm,LopesCardoso:2006bg,Cardoso:2008fr,Cardoso:2010gc,Cardoso:2014kwa}.
One could also try to adapt the entropy function formalism \cite{Sen:2005wa}, 
which employs universal properties of near horizon geometries and does not 
depend on supersymmetry.

Finally we comment on further possible future directions on the gravity 
side. Here it would be interesting to find solutions where other and
possibly more charges and fluxes have been turned on. We expect that our
formalism is particularly suited to finding dyonic solutions, due to
its built-in electric-magnetic covariance \cite{futurepaper}. 
For work in this direction it 
is encouraging that work on static BPS solutions in 
$U(1)$ gauged 
supergravity solutions with symmetric scalar target spaces 
has led to the construction of the general dyonic solution
\cite{Cacciatori:2009iz,Halmagyi:2013qoa,Katmadas:2014faa,Halmagyi:2014qza}.

We think that the systematic methods and explicit analytical
solutions interpolating between hvLif geometries that we have
presented in this paper 
will help to make progress towards a classification
of solutions in gauged supergravity, and of the hvLif landscape, and
to extend and deepen our understanding of the field theory/gravity
dictionary.

\section*{Acknowledgements}

We would like to thank G.~Lopes~Cardoso for drawing our attention to the
problem of constructing non-extremal versions of Nernst branes in 
gauged supergravity, and for many further useful discussions. We also
thank  P.~Athanasopoulos, M.~Haack, S.~Gentle, P.~Rakow and O.~Vaughan for useful discussions.
The work of PD is supported by STFC grant ST/J50113X/1 and National Research Foundation of Korea grants 2005-0093843, 2010-220-C00003 and 2012K2A1A9055280. 
The work of DE is supported by  STFC studentship ST/K502145/1.
The work of TM is partially supported by the STFC consolidated grants
ST/G00062X/1 and ST/L000431/1.

\appendix 

\section{Scalar potential in the real formulation of special geometry}\label{scalarpotapp}

In this appendix we review the real formulation of special geometry introduced
in \cite{Mohaupt:cmap},  based on the work of~\cite{Freed,Alekseevsky},
and extend it to include scalar potentials of
the form (\ref{scalpot}), which result from a flux superpotential
(\ref{superpot}). Starting from the holomorphic formulation,
where the complex scalars $X^I$ parametrise a conic affine special
K\"ahler (CASK) manifold, and where all vector multiplet couplings
are encoded in a holomorphic prepotential $F(X^I)$, which is homogeneous
of degree two, one introduces special real coordinates 
$(q^a)= \left( x^I, y_I \right)^T$, where
\[
X^I = x^I + i u^I, \quad F_I(X) = y_I + i v_I \;.
\]
Note that $F_I = \frac{\partial F}{\partial X^I}$ is homogeneous
of degree one. In the real formulation all vector multiplet
couplings are encoded in a Hesse potential $H(q^a)$, which 
is homogeneous of degree two, and which is obtained from the
imaginary part of the holomorphic prepotential by a Legendre 
transformation, which replaces $u^I$ by $y_I$ as an independent variable:
\begin{equation}\label{legendre}
H \left(x^I,y_I \right) = 2\, \mathrm{Im}\, F\! \left(X \left(x,y \right) \right) - 2 y_Iu^I \left( x,y \right) = \frac{i}{2} \left(X^I\bar{F}_I(X)-F_I(X)\bar{X}^I\right) \;.
\end{equation}
The special real coordinates $q^a$ are Darboux coordinates, and the
K\"ahler form on the CASK manifold is simply
\[
dx^I \wedge dy_I = \frac{1}{2} \Omega_{ab} dq^a \wedge dq^b \;,\;\;\;
\Omega_{ab} = \left( \begin{array}{cc}
0 & \id \\
- \id & 0 \\
\end{array} \right) .
\]

It is useful to note that the first derivatives $H_a$ of the
Hesse potential are related to the imaginary parts of $X^I$ and $F_I$
by
\[
H_a = 2 (v_I, - u^I)^T,
\]
and provide an alternative, `dual' coordinate system on the CASK manifold.

To obtain the associated projective special K\"ahler (PSK) manifold,
one imposes the D-gauge $-2H = \kappa^{-2}$, together with a condition
which fixes a $U(1)$ gauge. If one wants to preserve symplectic
covariance, one postpones fixing a $U(1)$ gauge and retains a local
$U(1)$ gauge invariance. Geometrically this corresponds to working
on the total space of a $U(1)$ principal bundle over the PSK manifold.

In \cite{Mohaupt:cmap} it was shown how to express all couplings appearing
in the bosonic part of the vector multiplet Lagrangian in terms
of real coordinates. In particular the CASK metric
$N_{IJ} = 2 \mbox{Im}F_{IJ}$ is replaced by the Hessian metric
\[
H_{ab} = \frac{\partial^2 H}{\partial q^a \partial q^b} \;.
\]

For the purpose of this paper we need to rewrite the scalar
potential $V(X,\bar{X})$ of (\ref{scalpot}), and the associated flux
superpotential $W(X)$ of (\ref{superpot}), in terms of real coordinates.
Using that $H_a = H_{ab}q^b$ by homogeneity, and using the formulae given above, it is straightforward to obtain
\begin{equation}\label{superpotqeqn}
W=W(q^a)=W(x^I,y_I)= 2g^a \left( \Omega_{ab} + \frac{i}{2} H_{ab} \right) q^b = i g^a \left( H_{ab} - 2 i \Omega_{ab} \right) q^b ,
\end{equation}
where we have defined $(g^a):=(g^I,g_I)^T$.

In order to obtain the potential $V$ as given in~\eqref{scalpot}, we must compute the derivatives 
\[
\partial_I W = \frac{\partial W}{\partial X^I} = \frac{1}{2} \left( \frac{\partial}{\partial x^I} - i\frac{\partial}{\partial u^I} \right) W.
\]
Since this derivative involves the real coordinates $(x^I,u^I)$ rather than $(q^a)=(x^I,y_I)^T$, we apply the chain rule to $W(x,y(x,u))$ and compute
\[
\left. \frac{\partial W}{\partial x^I} \right|_u = \left. \frac{\partial W}{\partial x^I} \right|_y + \left. \frac{\partial y_J}{\partial x^I} \frac{\partial W}{\partial y_J} \right|_x,
\quad
\text{and}
\quad \left. \frac{\partial W}{\partial u^I} \right|_x = \left. \frac{\partial y_J}{\partial u^I} \frac{\partial W}{\partial y_J} \right|_x .
\]
After decomposing the second derivatives of the prepotential $F$ into 
real and imaginary parts (including a conventional factor of 2)
by  $2F_{IJ}=R_{IJ}+iN_{IJ}$, one can apply the chain rule to show that 
\[ 
\frac{\partial y_J}{\partial x^I} = \frac{1}{2} \left( F_{IJ} + \bar{F}_{IJ} \right) = \frac{1}{2} R_{IJ},
\]
and read from~\cite{Mohaupt:cmap} that
\[
\frac{\partial y_J}{\partial u^I} = - \frac{1}{2} N_{IJ}.
\]
Combining this, we find 
\[
\frac{\partial W}{\partial X^I} = \frac{1}{2} \left( \frac{\partial W}{\partial x^I} + F_{IJ} \frac{\partial W}{\partial y_J} \right), \quad \frac{\partial \bar{W}}{\partial \bar{X}^I} = \frac{1}{2} \left( \frac{\partial \bar{W}}{\partial x^I} + \bar{F}_{IJ} \frac{\partial \bar{W}}{\partial y_J} \right).
\]
Finally, we can put all of this together to obtain
\begin{equation}\label{NderWeqn}
N^{IJ} \partial_I W \partial_J \bar{W} = \frac{1}{4} W_a \left( H^{ab} + \frac{i}{2} \Omega^{ab} \right) \bar{W}_b,
\end{equation}
where $\left( W_a \right) = \left( \frac{\partial W}{\partial x^I}, \frac{\partial W}{\partial y_J} \right)^T$, $H^{ab}$ is the inverse Hessian metric on the CASK manifold (see~\cite{Mohaupt:cmap}), and  $\Omega^{ab}$ is the inverse of $\Omega_{ab}$. 

Using~\eqref{superpotqeqn}, we have that
\begin{equation}\label{Wwithqeqn}
W_a = i g^b \left( H-2i \Omega \right)_{ba}, \qquad \bar{W}_a =-i g^b \left( H+2i \Omega \right)_{ba} .
\end{equation}
This can be substituted into~\eqref{NderWeqn}, which after simplification becomes
\begin{equation}\label{NderWwithqeqn}
N^{IJ} \partial_I W \partial_J \bar{W} =  H_{ab}g^a g^b,
\end{equation}
where we have used  the identity\footnote{This is the standard relation between the metric and K\"ahler form of a K\"ahler
manifold. The numerical factor is due to conventional choices.}  
$H_{ab} \Omega^{bc} H_{cd} = - 4 \Omega_{ad}$ \cite{Mohaupt:cmap}.

The final expression for the scalar potential given purely in terms of real coordinates then comes from~\eqref{scalpot} using~\eqref{superpotqeqn} and~\eqref{NderWwithqeqn} as follows:
\begin{align}\label{Vwithqeqn}
V &=  g^a H_{ab} g^b - 2\kappa^2 g^a \left( H_{ac} - 2i \Omega_{ac} \right) q^c g^b \left( H_{bd} + 2i \Omega_{bd} \right) q^d \notag
\\ &=  g^ag^b \left[ H_{ab} -2\kappa^2 H_aH_b -8 \kappa^2 \left( \Omega q \right)_a \left( \Omega q \right)_b \right] ,
\end{align}
where we have used homogeneity $H_a=H_{ab}q^b$.
Lastly, we substitute the $D$-gauge condition $-2H=\kappa^{-2}$  into \eqref{Vwithqeqn} to obtain
\begin{equation}
\label{V}
V = g^ag^b \left[ H_{ab} + \frac{H_aH_b + 4 \left( \Omega q \right)_a \left( \Omega q \right)_b}{H} \right].
\end{equation}
Note that the expression within the square brackets is homogeneous of
degree zero. This is useful in order to rewrite  $V$ in terms of
rescaled variables after dimensional reduction.

\section{Adapting the real formulation of special geometry to dimensional reduction}\label{appendix:realspecial}

We shall now define the various terms appearing in the three-dimensional Lagrangian~\eqref{3dLagr1}, which uses a modified version of the real formulation
of special geometry that is adapted to dimensional reduction. 
 We follow the conventions of~\cite{Mohaupt:cmap}, to which we refer the reader for further details. Firstly, the complex scalar fields, $X^I$, appearing in the four-dimensional Lagrangian~\eqref{4dLag}, are replaced by rescaled 
scalars
\begin{equation}\label{rescale}
Y^I:=e^{\phi /2} X^I\;,
\end{equation}
where $\phi$ is the Kaluza Klein scalar. In the four-dimensional 
theory parametrised by the $X^I$ the radial direction of the CASK
manifold, which is generated by the vector field
\[
\xi = X^I \frac{\partial}{\partial X^I} + \bar{X}^I \frac{\partial}{\partial 
\bar{X}^I} \, ,
\]
is a gauge degree of freedom. The above rescaling promotes it
to a physical degree of freedom, which is equivalent to the
Kaluza-Klein scalar. It turns out that this rescaling leads
to a convenient parametrization of the reduced three-dimensional
theory. Rewriting the D-gauge condition (\ref{dgaugex}) in terms of $Y^I$, we 
obtain 
\begin{equation}
-i \left( Y^I \bar{F}_I - F_I \bar{Y}^I \right) = e^\phi(Y, \bar{Y}),
\end{equation}
which determines $\phi$ in terms of the scalar fields $Y^I$.

Due to the homogeneity properties of the prepotential and Hesse potential, 
we can obtain a real parametrization which is based on the rescaled
complex scalars $Y^I$. The associated real coordinates are defined
by the decomposition
\[
Y^I=x^I+iu^I(x,y), \quad F_I(Y)=y_I + iv_I(x,y),
\]
as
\begin{equation}\label{qaofY}
q^a := \left( x^I , y_I \right)^T = \mathrm{Re} \left( Y^I,  F_I(Y) \right)^T.
\end{equation}

Furthermore, after reducing to three dimensions it is possible to write the gauge degrees of freedom using scalar fields as well. In particular, we define
\begin{equation}
\hat{q}^a:= \left( \frac{1}{2} \zeta^I , \frac{1}{2} \tilde{\zeta}_I \right)^T,
\end{equation}
where $\zeta^I$ are the components of the four-dimensional gauge fields $A_{\hat{\mu}}^I$ along the reduction direction, and $\tilde{\zeta}_I$ are the Hodge-duals of the three-dimensional vector parts. Specifically, these scalars descend from the four-dimensional field strengths as follows:
\begin{equation}\label{fieldstrengths}
\partial_\mu \zeta^I := F^I_{\mu 0}, \quad \partial_\mu \tilde{\zeta}_I := G_{I| \mu 0},
\end{equation}
where $G_{I| \hat{\mu} \hat{\nu}}$ are defined as
\[
G_{I|\hat{\mu}\hat{\nu}} := \mathcal{R}_{IJ}F^J_{|\hat{\mu}\hat{\nu}} -\mathcal{I}_{IJ}\tilde{F}^J_{\hat{\mu}\hat{\nu}}.
\]
 We can make further use of Hodge duality to encode the Kaluza-Klein vector degree of freedom using the scalar field $\tilde{\phi}$ \cite{Mohaupt:cmap}, although we will not need this here since we deal only with static configurations.

In terms of rescaled complex scalars $Y^I$ and rescaled real variables
$q^a$, the relation between prepotential $F(Y^I)$ and Hesse potential 
$H(q^a)$ is
\[
H \left(x^I,y_I \right) = 2\, \mathrm{Im}\, F\! \left(Y \left(x,y \right) \right) - 2 y_Iu^I \left( x,y \right) = \frac{i}{2} \left(Y^I\bar{F}_I(Y)-F_I(Y)\bar{Y}^I\right) = - \frac{1}{2} e^\phi.
\]
We also note that the D-gauge, when expressed in terms of rescaled real
scalars, reads
\begin{equation}\label{Dgaugespecre}
-2H \left( q^a \right) = e^\phi.
\end{equation}

In the Lagrangian~\eqref{3dLagr1}, we also use the tensor field
\begin{equation}\label{Htildedef}
\tilde{H}_{ab} := \frac{\partial^2}{\partial q^a \partial q^b} \tilde{H}, \quad \tilde{H}:= - \frac{1}{2} \log{ \left( - 2H \right) }.
\end{equation}
This tensor can be interpreted as a metric on the CASK manifold, which
is related to $H_{ab}$ by flipping the signature along the radial 
direction generated by the field $\xi$, combined with a conformal 
transformation which changes the scale transformation 
$q^a \rightarrow \lambda q^a$, where $\lambda \in \mathbbm{R}^{>0}$,
from being a homothety to being an isometry. This follows from the
obvious fact that while $H_{ab} dq^a dq^b$ is homogeneous of degree $2$,
$\tilde{H}_{ab} dq^a dq^b$ is homogeneous of degree $0$. Note that the
metric coefficients $H_{ab}$ and $\tilde{H}_{ab}$ are homogeneous of
degrees 0 and $-2$, respectively. Both tensors are related by
\begin{equation}
\label{HtildeH}
\tilde{H}_{ab} = \frac{1}{(-2H)} \left( H_{ab} - \frac{H_a H_b}{H} \right)\;.
\end{equation}

It will be convenient for us to introduce a set of dual coordinates with respect to the metric $\tilde{H}_{ab}$ defined by
\begin{equation}\label{H->q}
q_a:= \tilde{H}_a := \frac{\partial \tilde{H}}{\partial q^a} = -\frac{H_a}{2H}
= \frac{-1}{H} \left( \begin{array}{c}
v_I \\ - u^I \\
\end{array} \right) \;. 
\end{equation}
One can show that 
\begin{equation}
q_a=-\tilde{H}_{ab} q^b, \quad \partial_\mu q_a = \tilde{H}_{ab} \partial_\mu q^b,
\end{equation}
where we have used that $\tilde{H}_a$ is homogeneous of degree $-1$
for the first identity and the chain rule for the second.

It is also possible to use this metric to lower the index on $\partial_\mu\hat{q}^a$ to obtain the co-vector field
\begin{equation}
\partial_\mu \hat{q}_a:=\tilde{H}_{ab} \partial_\mu \hat{q}^b.
\end{equation}

Finally, we re-express the scalar potential in terms of variables 
adapted to dimensional reduction. Since, as we remarked, the
expression in the square brackets of (\ref{V}) is homogeneous of
degree zero, it remains invariant if we rescale the real 
coordinates $q^a$ by $e^{\phi/2}$. To express $V$ in terms
of the tensor $\tilde{H}_{ab}$ we use the relation 
\eqref{HtildeH}
to write
\begin{equation}
 V = - 2H g^ag^b \left[  \tilde{H}_{ab} -  \frac{H_aH_b}{H^2} - 2 \frac{\left( \Omega q \right)_a \left( \Omega q \right)_b}{H^2} \right].
\end{equation}
Finally, we use~\eqref{H->q} 
to re-write $V$ in terms of the dual coordinates $q_a$, and take into
account that upon dimensional reduction the term $-V$ in the Lagrangian
gets multiplet by $e^{-\phi} = -\frac{1}{2H}$ and obtain
\begin{equation}
\frac{1}{2H} V = - g^ag^b \left[\tilde{H}_{ab} - 4q_aq_b - 2 \frac{\left( \Omega q \right)_a \left( \Omega q \right)_b}{H^2} \right].
\end{equation}
\providecommand{\href}[2]{#2}\begingroup\raggedright\endgroup


\begin{thebibliography}{10}

\bibitem{Maldacena}
J.~M. Maldacena, {\it {The Large N limit of superconformal field theories and
  supergravity}},  {\em Int.J.Theor.Phys.} {\bf 38} (1999) 1113--1133,
  [\href{http://xxx.lanl.gov/abs/hep-th/9711200}{{\tt hep-th/9711200}}].

\bibitem{Hartnoll:2009sz}
S.~A. Hartnoll, {\it {Lectures on holographic methods for condensed matter
  physics}},  {\em Class.Quant.Grav.} {\bf 26} (2009) 224002,
  [\href{http://xxx.lanl.gov/abs/0903.3246}{{\tt arXiv:0903.3246}}].

\bibitem{Hartnoll:2011fn}
S.~A. Hartnoll, ``{Horizons, holography and condensed matter}.'' Chapter of the
  book `Black Holes in Higher Dimensions' to be published by Cambridge
  University Press (editor: G. Horowitz), 2011.

\bibitem{Dong}
X.~Dong, S.~Harrison, S.~Kachru, G.~Torroba, and H.~Wang, {\it {Aspects of
  holography for theories with hyperscaling violation}},  {\em JHEP} {\bf 1206}
  (2012) 041, [\href{http://xxx.lanl.gov/abs/1201.1905}{{\tt
  arXiv:1201.1905}}].

\bibitem{Sachdev:fermi}
L.~Huijse, S.~Sachdev, and B.~Swingle, {\it {Hidden Fermi surfaces in
  compressible states of gauge-gravity duality}},  {\em Phys.Rev.} {\bf B85}
  (2012) 035121, [\href{http://xxx.lanl.gov/abs/1112.0573}{{\tt
  arXiv:1112.0573}}].

\bibitem{Witten:1998qj}
E.~Witten, {\it {Anti-de Sitter space and holography}},  {\em
  Adv.Theor.Math.Phys.} {\bf 2} (1998) 253--291,
  [\href{http://xxx.lanl.gov/abs/hep-th/9802150}{{\tt hep-th/9802150}}].

\bibitem{Witten:1998zw}
E.~Witten, {\it {Anti-de Sitter space, thermal phase transition, and
  confinement in gauge theories}},  {\em Adv.Theor.Math.Phys.} {\bf 2} (1998)
  505--532, [\href{http://xxx.lanl.gov/abs/hep-th/9803131}{{\tt
  hep-th/9803131}}].

\bibitem{Chamblin:1999tk}
A.~Chamblin, R.~Emparan, C.~V. Johnson, and R.~C. Myers, {\it {Charged AdS
  black holes and catastrophic holography}},  {\em Phys.Rev.} {\bf D60} (1999)
  064018, [\href{http://xxx.lanl.gov/abs/hep-th/9902170}{{\tt
  hep-th/9902170}}].

\bibitem{landau2013statistical}
L.~Landau and E.~Lifshitz, {\em Statistical Physics}, vol.~5.
\newblock Elsevier Science, 2013.

\bibitem{Strominger:1996sh}
A.~Strominger and C.~Vafa, {\it {Microscopic Origin of the Bekenstein-Hawking
  Entropy}},  {\em Phys. Lett.} {\bf B379} (1996) 99--104,
  [\href{http://xxx.lanl.gov/abs/hep-th/9601029}{{\tt hep-th/9601029}}].

\bibitem{Maldacena:1997de}
J.~M. Maldacena, A.~Strominger, and E.~Witten, {\it {Black hole entropy in M
  theory}},  {\em JHEP} {\bf 9712} (1997) 002,
  [\href{http://xxx.lanl.gov/abs/hep-th/9711053}{{\tt hep-th/9711053}}].

\bibitem{Israel:1986}
W.~Israel, {\it Third law of black-hole dynamics: a formulation and proof},
  {\em Phys. Rev. Lett.} {\bf 57} (1986) 397--99.

\bibitem{Wald:1997qp}
R.~M. Wald, {\it {The `Nernst theorem' and black hole thermodynamics}},  {\em
  Phys.Rev.} {\bf D56} (1997) 6467--6474,
  [\href{http://xxx.lanl.gov/abs/gr-qc/9704008}{{\tt gr-qc/9704008}}].

\bibitem{DHoker:2009bc}
E.~D'Hoker and P.~Kraus, {\it {Charged Magnetic Brane Solutions in AdS$_5$ and
  the fate of the third law of thermodynamics}},  {\em JHEP} {\bf 1003} (2010)
  095.

\bibitem{Goldstein:2009cv}
K.~Goldstein, S.~Kachru, S.~Prakash, and S.~P. Trivedi, {\it {Holography of
  Charged Dilaton Black Holes}},  {\em JHEP} {\bf 1008} (2010) 078,
  [\href{http://xxx.lanl.gov/abs/0911.3586}{{\tt arXiv:0911.3586}}].

\bibitem{Goldstein:2010aw}
K.~Goldstein, N.~Iizuka, S.~Kachru, S.~Prakash, and S.~P. Trivedi, {\it
  {Holography of Dyonic Dilaton Black Branes}},  {\em JHEP} {\bf 1010} (2010)
  027, [\href{http://xxx.lanl.gov/abs/1007.2490}{{\tt arXiv:1007.2490}}].

\bibitem{Gauntlett:2009dn}
J.~P. Gauntlett, J.~Sonner, and T.~Wiseman, {\it {Holographic superconductivity
  in M-Theory}},  {\em Phys.Rev.Lett.} {\bf 103} (2009) 151601,
  [\href{http://xxx.lanl.gov/abs/0907.3796}{{\tt arXiv:0907.3796}}].

\bibitem{Horowitz:2009ij}
G.~T. Horowitz and M.~M. Roberts, {\it {Zero Temperature Limit of Holographic
  Superconductors}},  {\em JHEP} {\bf 0911} (2009) 015,
  [\href{http://xxx.lanl.gov/abs/0908.3677}{{\tt arXiv:0908.3677}}].

\bibitem{Barisch:2011ui}
S.~Barisch, G.~Lopes~Cardoso, M.~Haack, S.~Nampuri, and N.~A. Obers, {\it
  {Nernst branes in gauged supergravity}},  {\em JHEP} {\bf 1111} (2011) 090,
  [\href{http://xxx.lanl.gov/abs/1108.0296}{{\tt arXiv:1108.0296}}].

\bibitem{Horowitz:1997uc}
G.~T. Horowitz and S.~F. Ross, {\it {Naked black holes}},  {\em Phys.Rev.} {\bf
  D56} (1997) 2180--2187, [\href{http://xxx.lanl.gov/abs/hep-th/9704058}{{\tt
  hep-th/9704058}}].

\bibitem{LopesCardoso:1998wt}
G.~Lopes~Cardoso, B.~de~Wit, and T.~Mohaupt, {\it {Corrections to macroscopic
  supersymmetric black hole entropy}},  {\em Phys.Lett.} {\bf B451} (1999)
  309--316, [\href{http://xxx.lanl.gov/abs/hep-th/9812082}{{\tt
  hep-th/9812082}}].

\bibitem{LopesCardoso:2000qm}
G.~Lopes~Cardoso, B.~de~Wit, J.~Kappeli, and T.~Mohaupt, {\it {Stationary BPS
  solutions in N = 2 supergravity with $R^2$ interactions}},  {\em JHEP} {\bf
  12} (2000) 019, [\href{http://xxx.lanl.gov/abs/hep-th/0009234}{{\tt
  hep-th/0009234}}].

\bibitem{Dabholkar:2004dq}
A.~Dabholkar, R.~Kallosh, and A.~Maloney, {\it {A Stringy cloak for a classical
  singularity}},  {\em JHEP} {\bf 0412} (2004) 059,
  [\href{http://xxx.lanl.gov/abs/hep-th/0410076}{{\tt hep-th/0410076}}].

\bibitem{Sen:2005wa}
A.~Sen, {\it {Black hole entropy function and the attractor mechanism in higher
  derivative gravity}},  {\em JHEP} {\bf 0509} (2005) 038,
  [\href{http://xxx.lanl.gov/abs/hep-th/0506177}{{\tt hep-th/0506177}}].

\bibitem{heatup}
K.~Goldstein, S.~Nampuri, and {\'A}.~V{\'e}liz-Osorio, {\it {Heating up branes
  in gauged supergravity}},  {\em JHEP} {\bf 1408} (2014) 151,
  [\href{http://xxx.lanl.gov/abs/1406.2937}{{\tt arXiv:1406.2937}}].

\bibitem{BarischDick:2012gj}
S.~Barisch-Dick, G.~Lopes~Cardoso, M.~Haack, and S.~Nampuri, {\it {Extremal
  black brane solutions in five-dimensional gauged supergravity}},  {\em JHEP}
  {\bf 1302} (2013) 103, [\href{http://xxx.lanl.gov/abs/1211.0832}{{\tt
  arXiv:1211.0832}}].

\bibitem{Mohaupt:cmap}
T.~Mohaupt and O.~Vaughan, {\it {The Hesse potential, the c-map and black hole
  solutions}},  {\em JHEP} {\bf 1207} (2012) 163,
  [\href{http://xxx.lanl.gov/abs/1112.2876}{{\tt arXiv:1112.2876}}].

\bibitem{Klemm:2012yg}
D.~Klemm and O.~Vaughan, {\it {Nonextremal black holes in gauged supergravity
  and the real formulation of special geometry}},  {\em JHEP} {\bf 1301} (2013)
  053, [\href{http://xxx.lanl.gov/abs/1207.2679}{{\tt arXiv:1207.2679}}].

\bibitem{Klemm:2012vm}
D.~Klemm and O.~Vaughan, {\it {Nonextremal black holes in gauged supergravity
  and the real formulation of special geometry II}},  {\em Class.Quant.Grav.}
  {\bf 30} (2013) 065003, [\href{http://xxx.lanl.gov/abs/1211.1618}{{\tt
  arXiv:1211.1618}}].

\bibitem{Gnecchi:2013mja}
A.~Gnecchi, K.~Hristov, D.~Klemm, C.~Toldo, and O.~Vaughan, {\it {Rotating
  black holes in 4d gauged supergravity}},  {\em JHEP} {\bf 1401} (2014) 127,
  [\href{http://xxx.lanl.gov/abs/1311.1795}{{\tt arXiv:1311.1795}}].

\bibitem{staticaxfree}
D.~Errington, T.~Mohaupt, and O.~Vaughan, {\it {Non-extremal black hole
  solutions from the c-map}},  \href{http://xxx.lanl.gov/abs/1408.0923}{{\tt
  arXiv:1408.0923}}.

\bibitem{deWit:2001pz}
B.~de~Wit, {\it {Electric-magnetic duality in supergravity}},  {\em Nucl. Phys.
  Proc. Suppl.} {\bf 101} (2001) 154--171,
  [\href{http://xxx.lanl.gov/abs/hep-th/0103086}{{\tt hep-th/0103086}}].

\bibitem{Cardoso:2012nh}
G.~L. Cardoso, B.~de~Wit, and S.~Mahapatra, {\it {Non-holomorphic deformations
  of special geometry and their applications}},  {\em Springer Proc.Phys.} {\bf
  144} (2013) 1--58, [\href{http://xxx.lanl.gov/abs/1206.0577}{{\tt
  arXiv:1206.0577}}].

\bibitem{futurepaper}
P.~Dempster, D.~Errington, and T.~Mohaupt, {\em {\emph{ work in progress}}}.

\bibitem{Kallosh:2000rn}
R.~Kallosh, T.~Mohaupt, and M.~Shmakova, {\it {Excision of singularities by
  stringy domain walls}},  {\em J.Math.Phys.} {\bf 42} (2001) 3071--3081,
  [\href{http://xxx.lanl.gov/abs/hep-th/0010271}{{\tt hep-th/0010271}}].

\bibitem{Mayer:2003zk}
C.~Mayer and T.~Mohaupt, {\it {The K\"ahler cone as cosmic censor}},  {\em
  Class. Quant. Grav.} {\bf 21} (2004) 1879--1896,
  [\href{http://xxx.lanl.gov/abs/hep-th/0312008}{{\tt hep-th/0312008}}].

\bibitem{perlmutter}
E.~Perlmutter, {\it {Hyperscaling violation from supergravity}},  {\em JHEP}
  {\bf 1206} (2012) 165, [\href{http://xxx.lanl.gov/abs/1205.0242}{{\tt
  arXiv:1205.0242}}].

\bibitem{Bueno:2012sd}
P.~Bueno, W.~Chemissany, P.~Meessen, T.~Ortin, and C.~Shahbazi, {\it
  {Lifshitz-like Solutions with Hyperscaling Violation in Ungauged
  Supergravity}},  {\em JHEP} {\bf 1301} (2013) 189,
  [\href{http://xxx.lanl.gov/abs/1209.4047}{{\tt arXiv:1209.4047}}].

\bibitem{Bueno:2012vx}
P.~Bueno, W.~Chemissany, and C.~Shahbazi, {\it {On $hvLif$-like solutions in
  gauged Supergravity}},  {\em Eur.Phys.J.} {\bf C74} (2014), no.~1 2684,
  [\href{http://xxx.lanl.gov/abs/1212.4826}{{\tt arXiv:1212.4826}}].

\bibitem{Mayer:2004sd}
C.~Mayer and T.~Mohaupt, {\it {Domain walls, Hitchin's flow equations and
  G(2)-manifolds}},  {\em Class.Quant.Grav.} {\bf 22} (2005) 379--392,
  [\href{http://xxx.lanl.gov/abs/hep-th/0407198}{{\tt hep-th/0407198}}].

\bibitem{futurepaper5d}
P.~Dempster, D.~Errington, and T.~Mohaupt, {\em {\emph{to appear}}}.

\bibitem{LopesCardoso:2006bg}
G.~Lopes~Cardoso, B.~de~Wit, J.~Kappeli, and T.~Mohaupt, {\it {Black hole
  partition functions and duality}},  {\em JHEP} {\bf 0603} (2006) 074,
  [\href{http://xxx.lanl.gov/abs/hep-th/0601108}{{\tt hep-th/0601108}}].

\bibitem{Cardoso:2008fr}
G.~L. Cardoso, B.~de~Wit, and S.~Mahapatra, {\it {Subleading and
  non-holomorphic corrections to N=2 BPS black hole entropy}},
  \href{http://xxx.lanl.gov/abs/0808.2627}{{\tt arXiv:0808.2627}}.

\bibitem{Cardoso:2010gc}
G.~L. Cardoso, B.~de~Wit, and S.~Mahapatra, {\it {BPS black holes, the Hesse
  potential, and the topological string}},  {\em JHEP} {\bf 06} (2010) 052,
  [\href{http://xxx.lanl.gov/abs/1003.1970}{{\tt arXiv:1003.1970}}].

\bibitem{Cardoso:2014kwa}
G.~Cardoso, B.~de~Wit, and S.~Mahapatra, {\it {Deformations of special
  geometry: in search of the topological string}},
  \href{http://xxx.lanl.gov/abs/1406.5478}{{\tt arXiv:1406.5478}}.

\bibitem{Cacciatori:2009iz}
S.~L. Cacciatori and D.~Klemm, {\it {Supersymmetric AdS(4) black holes and
  attractors}},  {\em JHEP} {\bf 1001} (2010) 085,
  [\href{http://xxx.lanl.gov/abs/0911.4926}{{\tt arXiv:0911.4926}}].

\bibitem{Halmagyi:2013qoa}
N.~Halmagyi, {\it {BPS Black Hole Horizons in N=2 Gauged Supergravity}},  {\em
  JHEP} {\bf 1402} (2014) 051, [\href{http://xxx.lanl.gov/abs/1308.1439}{{\tt
  arXiv:1308.1439}}].

\bibitem{Katmadas:2014faa}
S.~Katmadas, {\it {Static BPS black holes in U(1) gauged supergravity}},
  \href{http://xxx.lanl.gov/abs/1405.4901}{{\tt arXiv:1405.4901}}.

\bibitem{Halmagyi:2014qza}
N.~Halmagyi, {\it {Static BPS Black Holes in AdS4 with General Dyonic
  Charges}},  \href{http://xxx.lanl.gov/abs/1408.2831}{{\tt arXiv:1408.2831}}.

\bibitem{Freed}
D.~S. Freed, {\it {Special Kahler manifolds}},  {\em Commun.Math.Phys.} {\bf
  203} (1999) 31--52, [\href{http://xxx.lanl.gov/abs/hep-th/9712042}{{\tt
  hep-th/9712042}}].

\bibitem{Alekseevsky}
D.~Alekseevsky, V.~Cortes, and C.~Devchand, {\it {Special complex manifolds}},
  {\em J.Geom.Phys.} {\bf 42} (2002) 85--105,
  [\href{http://xxx.lanl.gov/abs/math/9910091}{{\tt math/9910091}}].

\end{thebibliography}
\end{document}